\documentclass[usenatbib]{mn2e}
\usepackage{epsfig,lscape}

\newcommand{\etal}{et~al.} 

\newcommand{\UCHII}{UCH{\sc ii} }

\newcommand{\kms}{$\mbox{km~s}^{-1}$ }
\newcommand{\kmsns}{$\mbox{km~s}^{-1}$}

\newcommand{\vsfig}[2]           % Single FIGure (put one figures in the
{
  \begin{center}
    \begin{minipage}[t]{0.05\textwidth}
      {\footnotesize \raisebox{40mm}{(#2)}}
    \end{minipage}
    \begin{minipage}[t]{0.42\textwidth}
      \psfig{file=./#1.ps,height=0.95\textwidth,angle=0}
    \end{minipage}
    \hfill
  \end{center}
}

\newcommand{\specdfig}[2]        % Double FIGures (put two figures  
{
   \begin{center}
     \begin{minipage}[t]{0.45\textwidth}
         \psfig{file=eps/#1.eps,height=0.65\textwidth,width=1\textwidth,angle=0}
     \end{minipage}
     \hfill
     \begin{minipage}[t]{0.45\textwidth}
         \psfig{file=eps/#2.eps,height=0.65\textwidth,width=1\textwidth,angle=0}
     \end{minipage}
   \end{center}
}

\newcommand{\specsfig}[1]        % Single FIGure (put one figure of  
{
   \begin{center}
     \begin{minipage}[t]{0.45\textwidth}
         \psfig{file=eps/#1.eps,height=0.65\textwidth,width=1\textwidth,angle=0}
     \end{minipage}
   \end{center}
}

\newcommand{\boxfig}[1]        % Single FIGure (put one figure of  
{
   \begin{center}
     \begin{minipage}[t]{0.46\textwidth}
         \psfig{file=eps/#1.eps,height=0.45\textwidth,angle=0}
     \end{minipage}
   \end{center}
}

\newcommand{\twofig}[2]        % Double FIGures (put two figures  
{
   \begin{center}
     \begin{minipage}[t]{0.5\textwidth}
         \psfig{file=eps/#1.eps,height=0.95\textwidth}
     \end{minipage}
     \hfill
     \begin{minipage}[t]{0.5\textwidth}
         \psfig{file=eps/#2.eps,height=0.95\textwidth}
     \end{minipage}
   \end{center}
}

\begin{document}

\title[Water masers towards 1.2-mm dust clumps]{Constraining the properties of 1.2-mm dust clumps that contain luminous water masers}
\author[S.\ L.\ Breen \& S.\ P. Ellingsen]{S.\ L. Breen$^{1,2}$\thanks{Email: Shari.Breen@csiro.au} and S.\ P. Ellingsen$^2$ \\
  \\
  $^1$ CSIRO Astronomy and Space Science, PO Box 76, Epping, NSW 1710, Australia\\
  $^2$ School of Mathematics and Physics, University of Tasmania, Private Bag 37, Hobart, Tasmania 7001, Australia\\
}
 
 \maketitle
  
 \begin{abstract}
 We have conducted a sensitive water maser search with the Australia Telescope Compact Array towards 267 1.2-mm dust clumps presented in the literature. We combine our new observations with previous water maser observations to extend our sample to 294 1.2-mm dust clumps, towards which we detect 165 distinct water maser sites towards 128 1.2-mm dust clumps.  Within the fields of our observations, we additionally find four water masers with no apparent associated 1.2-mm dust continuum emission. Our overall detection rate of 44 per cent appears to vary as a function of Galactic longitude. We find that there is an excellent correspondence between the locations of the detected water masers with the peak of the target 1.2-mm dust clump sources. As expected from previous similar studies, the water masers are chiefly detected towards the bigger, brighter and more massive 1.2-mm dust clumps. 
 
We find further evidence to suggest that the water masers tend to increase in flux density (and therefore luminosity), as well as velocity range, as the sources evolve. We also show that the current sample of water maser sources suffer less from evolutionary biases than previous targeted searches. 

We have compared the locations of the water masers with dust clumps which have a previously determined association with 6.7-GHz methanol masers and 8-GHz radio continuum. We find that a higher fraction of 1.2-mm dust clump sources in our sample are only associated with water masers (41) than only associated with methanol masers (13). This suggests that water masers can be present at an even earlier evolutionary stage than 6.7-GHz methanol masers. Comparison of the water maser detection rates associated with different combinations of methanol maser and radio continuum, as well as those with neither tracer, shows that the highest detection rate is towards those sources which also exhibit methanol maser emission.

We have tested a previously hypothesised model for water maser presence towards 1.2-mm dust clumps. We find water masers towards a high proportion of the clumps that the model predicts would have associated water masers but also find a number of water masers towards sources with a low calculated probability. We propose that this is likely an artifact of the poorly determined distances to the sources. We suggest refinements and future work which will further constrain the nature of the driving sources associated with water masers.

%
%It is apparent that the water maser detection rate differs for dust clumps in different association categories 
% 
% 
%Comparison between our maser sources with other large samples shows that the average velocity range and peak flux density of sources are lower than for other samples. Given that our data are in support of a scenario whereby the water masers increase in flux density and velocity range as the sources evolve, combined with the tendency for the comparison searches to chiefly target more evolved sources, leads us to the conclusion that our current sample of sources  have been detected towards sources across a broader range of evolutionary stages. 
%
%Association rates with meth cont combinations

\end{abstract}


\section{Introduction}

To date there have been relatively few large-scale, unbiased surveys
for water masers. Instead, searches have focused on high-mass star formation regions, typically targeting regions selected on the basis of {\em IRAS} colours
\citep[e.g.][]{Churchwell1990,Codella95,Sunada07} or other maser species \citep[e.g.][]{B80,C+83,Beuther02}. Due
to this, the known sample of water masers lacks the completeness that
has been achieved for both 6.7-GHz methanol masers \citep[e.g.][]{CasMMB10, GreenMMB10} and main-line OH masers \citep[e.g.][]{C98}. This,
coupled with the fact that the positions of the majority of southern
hemisphere water masers are not known to sub-arcsecond accuracy, means
that relatively little is known about the nature of sources that are able to produce luminous water masers.

% Targeted searches for 22 GHz water masers are therefore popular, although these have been traditionally biased towards more evolved sources (as traced by e.g. {\em IRAS} sources \citep[e.g.][]{Churchwell1990} or other maser species \citep[e.g.][]{B80,C+83,Beuther02}).

Large-scale, unbiased surveys for water masers are hindered by the large amounts of time required, and the susceptibility of their observation to poor weather conditions. Further complications arise from the intrinsic variability of the sources which requires high positional precision in either the initial observations, or rapid follow up observations. Some systematic searches for water masers have been attempted \citep[e.g.][]{Matthews85,Breen,CB10,CBE10} but generally cover only very small regions of the Galactic plane. However, the observations of a relatively shallow, but large-scale single-dish survey for water masers have recently been completed, covering 100 degrees in longitude of the Southern Galactic Plane \citep{Walsh08} and the survey results are expected in the coming year.

Due to the difficulties associated with complete searches, attempts have been made to model water maser presence towards high-mass star formation regions, with the aim of gaining a more complete sample of water maser sources. These attempts have used particular observable properties of the high-mass star formation regions to make these predictions. \citet{Palla91} modelled the probability of water maser presence on the
basis of {\em IRAS} far-infrared colours.  This model was tested by
\citet{Palla93} and resulted in a detection rate of around
5 per cent.  This disappointing result is most likely due to the poor spatial
resolution of the {\em IRAS} observations (30 arcseconds at
100~$\mu$m) which results in significant source confusion in crowded
high-mass star formation regions; demonstrating the need for
high-resolution complementary data in the formulation of such models.

A systematic search for water masers of almost half a square degree
of the G\,333.2--0.6 giant molecular cloud \citep{Breen} revealed
a strong relationship between water maser presence and the bigger,
brighter, more massive and dense 1.2-mm dust clumps (identified by
\citet{Mook04}), within the giant molecular cloud.  From these
observations, \citet{Breen} produced what appears to be a
reliable statistical model for predicting which 1.2-mm dust clumps
will and will not have an associated water maser. While the likelihood of water maser presence increases with increasing values of all dust clump properties tested, the simplest model with the greatest predictive properties was found to only include the dust clump radius. This model predicts that all 1.2-mm dust clumps of radius 1.25~pc (or higher) have a probability of 0.5 (or higher) of having an associated water maser.

While the model presented in \citet{Breen} is promising, the water maser survey observations on which this model was derived, were of comparatively low sensitivity (detection limit of $\sim$5~Jy or more in some regions). This, combined with the fact that the sample size was small, means that its validity cannot be determined before testing and refining it on a much larger sample. A reliable model relating the properties of 1.2-mm dust clumps to the presence/absence of water masers represents a unique tool for investigating questions such as the mass range of the stars that produce luminous water masers and the evolutionary phase they trace. Here we perform the necessary observations to test and refine the model developed in \citet{Breen}. 

 \citet{Hill05} present a catalogue of 404 1.2-mm dust clumps, a large but manageable sample, perfect for a targeted water maser search to test the current model. \citet{Hill05} targeted their 1.2-mm dust continuum
observations towards 131 regions that were suspected of undergoing
massive star formation, using the presence of previously identified
methanol masers and/or \UCHII regions to select
their targets. The observations were carried out using the SIMBA instrument on SEST (the 15~m Swedish ESO Submillimetre telescope) and detected emission directly associated with all but 20 of the methanol masers and nine of the \UCHII regions targeted. \citet{Hill05} also made a large number of serendipitous detections within the target fields. 

Here we present new Australia Telescope Compact Array (ATCA) observations towards 267 of the \citet{Hill05} dust clumps and supplement our observations with data taken from the literature \citep{Breen10b,CB10,FC89}. Altogether we present water maser data towards 294 dust clump. Using the model of \citet{Breen} we find that few of these dust clumps have high calculated probabilities of water maser presence (only 58 have probabilities greater than 0.01).

\citet{Breen10a} present 12.2-GHz methanol maser observations targeting a number of these dust sources. This search for 12.2-GHz methanol masers, constitutes a near-complete sample within the sample of \citet{Hill05} sources, since no strong 12.2-GHz emission is expected in regions lacking 6.7-GHz methanol maser emission. Since a number of the sources presented in \citet{Hill05} are contained within the regions searched for OH maser emission by \citet{C98}, a large subset of the 1.2-mm dust clumps have been searched for OH masers. %Here, new 22-GHz water maser observations completed with the ATCA towards much of the \citet{Hill05} sample are presented. 

%In some instances, data presented in \citet{Breen10b} are used, rather that carrying out repeat observations and likewise observations of \citet{FC89} and \citet{CB10} are used wherever possible. However, since many of the 1.2-mm dust clumps are located near other dust clumps, even in the case where one of those dust clumps is associated with a previously reported water maser, they were often observed again since they were located within the field of nearby dust clumps. 

%We find that there are approximately 100 1.2-mm dust
%clumps in the catalogue of Hill et al. for which our model gives a
%probability of 0.1 or higher of water maser presence.  We have used
%the University of Tasmania 26m radio telescope to search all 404 1.2
%mm dust clumps and have detected a total of 240 water masers. Given
%the limited pointing accuracy of a 26m radio telescope (approximately
%0.5 arcminutes for Mt Pleasant) we are not able to determine if the
%detected water masers are truely associated with the targeted 1.2-mm
%dust clumps or merely a chance detection.

\section{Observations and data reduction}\label{sect:obs}

\subsection{Observations}

The feasibility of this study was tested with observations using the University of Tasmania's 26 m Mount Pleasant radio telescope. All 404 1.2-mm dust clumps given in \citet{Hill05} were targeted in these observations during 2007. An estimated 150 water maser sources were detected within the regions. An effort was made to locate the origin of the emissions by observing four-point grids around each of the targets and fitting a 2D-Gaussian to the relative amplitudes of the emission detected in each. Given the immense success in detecting water maser sources, together with the impossible task of confidently disentangling nearby sources and associating them with the 1.2-mm dust clumps (especially given the limited pointing accuracy of the instrument), we proposed to observe the sources with the ATCA. Using the ATCA has several advantages, foremost of which are the ability to obtain accurate positions and gain much greater sensitivity.

\begin{table}
\begin{center}
  \caption{ATCA observations: epochs, array configurations and typical synthesised beam sizes.}
\begin{tabular}{|l|l|l}\hline
 {\bf Array}  & \multicolumn{1}{c}{\bf Epoch} &{\bf Beam size} \\ \hline
H75	&  2007 October 9, 10 & 28x24   \\
%6D	&  2007 November 26  \\
H214	&	2007 December 7 &  13x8\\
H214	&  2008 July 12, 13 &  13x8 \\
6B	&  2008 August 15, 16, 17 &1.7x0.5  \\ \hline

	\end{tabular}\label{tab:array}
\end{center}
\end{table}

%\begin{table}
%\begin{center}
%  \caption{Typical size of the synthesised beam in each of the array configurations.}
%\begin{tabular}{|l|l|}\hline
% {\bf array}  &{\bf beam size}  \\ 
% &{\bf (arcsec)}\\ \hline
%H75	&  28x24  \\
%H214	&  13x8 \\
%6B	&  1.7x0.5 \\ \hline
%	\end{tabular}\label{tab:beam}
%\end{center}
%\end{table}

Successful ATCA observations were carried out in three different array configurations over four epochs; these are listed in Table~\ref{tab:array}. For 137 of the 404 1.2-mm continuum sources observed by \citet{Hill05}, observations at 22-GHz are not presented. The reasons for the omission of these 137 sources were chiefly due to their: proximity to a declination of zero degrees (where observations can be particularly troubling at the ATCA); location in the northern hemisphere; lack of reported dust characteristics; or because the weather conditions they were taken in were too poor to result in reliable data. Observations of an additional 27 1.2-mm dust clumps were not necessary since they are associated with water masers reported to good positional accuracy in the literature.

In total, we successfully searched 267 1.2-mm dust clumps for water maser presence with the ATCA. For all observations the field of view of the telescope is almost 5 arcmin, and the full width half maximum of the primary beam is 2.3 arcmin. Observations were carried out in groups of sources in the same neighbourhood so as to decrease overheads and allowing reference pointing observations to be carried out efficiently. Pointings were carried out on a strong continuum source, located within $\sim$20 degrees of the target sources, approximately every hour. Using this method, the pointing is usually accurate to 5 arcsec, compared to about 20 arcsec when no pointing observations are carried out.

Targets were observed in a series of cuts, interspersed with phase calibration observations. At all epochs of observations only a single linearly polarised signal was recorded. The general observation strategy differed depending on the array and weather conditions, but in general, nearby sources were grouped together and bracketed by phase calibration observations carried out on a compact continuum source, located not more than 10 degrees from the targets. The time spent observing the targets and the time between phase calibration was tailored to most efficiently observe the targets in the allocated time on the ATCA. The average strategy was to observe each target field for 10 minutes, over five cuts, with phase calibration (onsource for 1.5 minutes) carried out every 10 minutes. In general, these observations have allowed 5-$\sigma$ detections to be made at the 150 - 200~mJy level in the central section of the beam. For some sources, integration times were adjusted depending on the scheduled time allotment, and due to this the sensitivity limit may be either slightly lower or slightly higher (these are listed in column 10 of Table~\ref{tab:dust_sources}).

On each observing day, observations of a bandpass calibrator were completed. Two different sources were used; PKS B1253--055 and PKS B1921--293, both are strong at 22~GHz and are usefully observable at high elevations near the beginning and end of the time that the Galactic plane is visible, respectively. Primary flux calibration is with respect to observations of PKS B1934--638, which were also carried out daily. 

Details of the the observations carried out in each of the array configurations are given below. Descriptions of the correlator configurations (and therefore spectral resolution and velocity coverage) and individual observation strategies are also given in this section. %Note that none of the data were smoothed.

\subsubsection{H75}\label{sect:h75}

The first series of observations were carried out in the most compact ATCA configuration, H75. Primarily, observations of sources located near declination zero were intended for this array configuration, but we were able to observe some additional, early rising sources in a Director's time allocation prior to our scheduled time on 2007 October 9.

For these observations the correlator was configured to record 1024 channels across a bandwidth of 16 MHz for a single polarisation. This correlator configuration allowed for a channel spacing of 0.21 \kmsns, corresponding to a velocity resolution of 0.25 \kmsns. The velocity coverage of these observations was just over 200 \kmsns.   

In these observations, target sources were observed for 10 minutes each, over a series of five cuts (in general) spread over several hours. Observations of a phase calibrator source were carried out every 8 - 14 minutes.

\subsubsection{H214}\label{sect:h214}

Water maser observations were carried out in the H214 hybrid configuration over two epochs. The first epoch was during 2007 December 7 and included only one source (G\,213.705--12.597) which was observed as part of a multifrequency study of the masers associated with nearby star formation region Mon R2. The adopted correlator configuration for these observations were very different for this source, using a bandwidth of 4 MHz and 1024 channels. In this configuration the velocity coverage is limited to $\sim$50 \kmsns, but the spectral resolution is increased to 0.06 \kmsns.

The main series of observations carried out in this array configuration were completed in 2008 July. For these observations the correlator was configured to record 512 channels over a bandwidth of 32 MHz for a single polarisation. We made this change in correlator configuration, as from our first series of observations, it was clear that for some sources we were unable to observe the full extent of the velocity range of the emission. In these observations we made a sensible compromise, forgoing additional spectral resolution for twice the bandwidth. The resultant velocity coverage is over 400 \kms and the spectral resolution is 1.0 \kmsns. The average strategy for these observations was to observe each source for 12-14 minutes over 6-7 cuts of 2 minutes each spaced over several hours. Phase calibrator observations were completed approximately every 13 minutes.

%Although scheduled with the intention of observing only those sources near a declination of zero degrees, a number of hours of Director's time were awarded on 2008 July 12 and 13 from LST 6h. This allowed a number of the dust sources with right ascensions chiefly less than 13h to additionally be observed.

%\begin{table}
%\begin{center}
%  \caption[H214 configuration antenna spacings]{H214 configuration. Spacings between each pair of the antennas are given in metres.}\vspace{0.5cm}
%\begin{tabular}{|c|c|c|c|c|c|}\hline
% 	&{\bf CA02} &	{\bf CA03}	 & {\bf CA04} &	{\bf CA05} & 	{\bf CA06}\\ \hline
%{\bf CA01}	&	 92	& 230&	 144	& 247	& 4500\\
%{\bf CA02}&	  	 &138 & 	 82& 	 216	& 4408\\
%{\bf CA03}& 	  	 &&	 132	 & 240	 &4270\\
%{\bf CA04}&	  	 &&& 	  	 138	& 4378\\
%{\bf CA05}&	  	  	  	 &&&& 	 4383\\ \hline
%	\end{tabular}\label{tab:h214}
%\end{center}
%\end{table}

\subsubsection{6B}\label{sect:6b}

The final series of observations was carried out in a 6B array in 2008 August. The correlator was configured as for the observations carried out in the H214 array, giving a velocity coverage of more than 400 \kmsns, and a spectral resolution of 1.0 \kmsns. Similar to the other observations, the observing strategy for observations in this array saw sources observed as a series of 2 minute cuts spaced over several hours. On average, sources were observed seven times, giving onsource integration times of 14 minutes. Observations of a phase calibrator were conducted approximately every seven minutes.

%These observations included a large proportion of the dust clump sources that were targeted, as the time allocated on the ATCA was quite large and also included a number of hours of Director's time on either side of the scheduled time on each day. Due to this, we were able to split the time up into essentially two observing sessions per day, filling the first with early rising sources, and the second with the later rising sources, and were still able to achieve sufficient {\em uv}-coverage to confidently position any detected sources.

%
%\begin{table}
%\begin{center}
%  \caption[6B configuration antenna spacings]{6B configuration. Spacings between each pair of the antennas are given in metres.}\vspace{0.5cm}
%\begin{tabular}{|c|c|c|c|c|c|}\hline
% 	&{\bf CA02} &	{\bf CA03}	 & {\bf CA04} &	{\bf CA05} & 	{\bf CA06}\\ \hline
%{\bf CA01}	&	 949 &	 2219	&  2755	&  2969&	 5969\\
%{\bf CA02}&	  	 &1270	& 1806	& 2020&	 5020\\
%{\bf CA03}& 	  	 &&	 536	&  750&	 3750\\
%{\bf CA04}&	  	 &&& 	  	  214	 &3214\\
%{\bf CA05}&	  	  	  	 &&&& 	 3000\\ \hline
%
%	\end{tabular}\label{tab:6b}
%\end{center}
%\end{table}
%

\subsection{Data reduction}

All ATCA data were reduced using the {\sc miriad} software package \citep{Sault}, applying the standard techniques for ATCA spectral line observations. Image cubes of the entire primary beam and usable velocity range were produced for each source. The flux densities 
of sources that were located away from the centre of the primary beam have been 
corrected to account for beam attenuation using the {\sc miriad} task `linmos'. Both vector and scalar average spectra of the calibrated {\em uv}-data were inspected for each of the targets and cross checked with the emission identified in each of the image cubes. Spectra for each of the detected sources were produced by integrating the emission in the image cubes. The typical resultant rms noise in each spectrum was 25 - 40 mJy for sources located near the centre of the beam.

The typical synthesised beam sizes experienced in each of the array configurations are listed in Table~\ref{tab:array}. In both of the hybrid configurations (H75 and H214), data from the 6 km antenna (antenna 6) was not included in the data reduction. The adopted water maser rest frequency was 22235.07985~MHz.

\section{Results}

\citet{Hill05} presented a catalogue of 404 1.2-mm dust clumps. Towards 294 of these clumps, we present water maser data and compare our detections with the derived probability of water maser presence (using the model presented in \citet{Breen}). Most of the water maser data presented are new observations (132 of 165 water masers), however, where appropriate, data from \citet{Breen10b}, \citet{CB10} and \citet{FC89} has been used to minimise repetition of observations.

Table~\ref{tab:dust_sources} shows the targeted 1.2-mm dust clumps, along with their calculated probability of water maser presence, followed by the water maser data. We find a total of 165 water maser sources towards 128 of the 1.2-mm dust clumps. Spectra of each of the 132 water masers detected in our current observations are presented in Fig.~\ref{fig:dust_water_spectra} and are essentially ordered as the 1.2-mm dust clumps presented in \citet{Hill05} (in order of increasing right ascension) unless there was a need for nearby sources to be vertically aligned. For the majority of sources, a velocity range of 200 \kms is shown, however, this has been reduced to 100 \kms for sources observed in the H75 configuration of the ATCA (which were carried out with a smaller bandwidth) and for one source,  G\,213.705--12.597, a velocity range of 40 \kms is shown (this source was observed with 4 MHz bandwidth). Confusion from strong nearby sources are marked on the individual spectra, except where they are present as features of negative intensities. In some spectra, where confusing features from multiple nearby sources are present, we use an `X' to indicate that those features are not associated with the presented source. A consequence of the relatively coarse spectral resolution of our observations, combined with the often strong and relatively narrow water maser features, is that `ringing' is present in a number of spectra. 

A number of the sources that we detect have been previously presented in the literature \citep[e.g.][and references therein]{John72,C74,K76,GD77,B80,BS82,BE83,C+89,HC96}. However, the majority of these observations were conducted 20 or more years ago, with relatively poor positional accuracy. When combined with the often extreme variable nature of water maser sources, the task of identifying our sources with those in the literature is therefore fraught with difficulties. Since the only certain way to match up sources with those in the literature is by position, we only compare our detections with observations of comparable positional accuracy. 

Several of the sources that we detect have been observed in interferometric observations \citep[e.g.][]{Breen10b,CB10,CP08,FC89}, allowing us to compare our positions and therefore assessing our positional uncertainties. In general, our positional agreement with previous observations is excellent, sometimes better than 0.5 arcsec, but consistently within 2 arcsec. \citet{Breen10b} present a more extensive assessment of the positional uncertainties of their water maser observations carried out with the ATCA  and conclude that perhaps the largest contributor to the uncertainty was the tendency for the water maser spots to be spread out and the difference between measuring a slightly different feature at a different epoch. Here we similarly find that this is the case, and therefore expect the positional uncertainties of the water masers that we present to be accurate to 2 arcsec.

Another facet of this assessment allowed us to determine the effect that the synthesised beam has on the positional accuracy of the sources. As can be seen in Table~\ref{tab:array}, the beam sizes resulting from the hybrid configurations are much larger than that of the more extended 6B array. Comparisons with other measurements shows that even when the beam is large, accurate positions can be derived, since it is rarely the case that two sources show emission at the same velocities and are spatially separated by less than a synthesised beam. The much smaller beam of the 6B array has allowed for some water maser sites to be broken up into individual components, although it is likely the case that very close companions are intimately associated with the same exciting source.

Notes on some individual sources are presented in Section~\ref{sect:water_dust_ind}. We find that there is an excellent correspondence between the locations of the detected water masers and the 1.2-mm dust sources and this is discussed further in Section~\ref{sect:water_dust}. In addition to the water masers that we detect towards 1.2-mm dust clumps, we detect four water maser sources that seem not to be associated with any dust continuum emission and these are discussed in Section~\ref{sect:special}. 

\begin{landscape}
\begin{table}

\begin{center}
  \caption[Target 1.2-mm dust clumps, probability of water maser association, followed by a description of the water maser observations including any detections.]{Target 1.2-mm dust clumps, probability of water maser association, followed by a description of the water maser observations including any detections. Columns 1-3 give the 1.2-mm dust clump name followed by the right ascension and declination \citep{Hill05} and column 4 shows the calculated probability of water maser presence (using the model derived by \citet{Breen}). Column 5 gives information about the ATCA array configuration used in the water maser observations followed by the epoch. In some cases, data is taken from \citet{Breen10b} which is marked in column 5 with a `BCEP' followed by the year of the observations. This format is also followed for sources taken from \citet{CB10} (CB) and \citet{FC89} (FC). Columns 6-13 give details of the water masers detected towards the dust clumps (which in some cases is more than one source), as well as 5-$\sigma$ detection limits where no source is detected, specifically; column 6: water maser Galactic coordinates; column 7: water maser right ascension; column 8: water maser declination; column 9: separation between water maser and 1.2-mm dust clump peak (arcsec); column 10: water maser peak flux density (Jy) or if preceded by `$<$' the 5-$\sigma$ detection limit (mJy); column 11: velocity of the water maser peak emission (\kmsns); column 12: velocity range of either the detected emission, or the observed velocity range if no detection; and column 13: integrated flux density of the water maser emission (Jy \kmsns). Column 14 presents notes related to the sources, which in most cases are further discussed in Section~\ref{sect:water_dust_ind} (note the following abbreviations: BS(RS) - indicates that the source is dominated by a blue(red)-shifted feature.)}
% \vspace{0.5cm}
 \footnotesize
\begin{tabular}{lccrllcccllcrl}\hline
    \multicolumn{3}{c}{\bf Dust clump} & &  \multicolumn{10}{c}{\bf Water maser} \\
 \multicolumn{1}{c}{\bf Name} & {\bf RA}  & {\bf Dec} & \multicolumn{1}{c}{\bf prob.} & \multicolumn{1}{c}{\bf array} &  \multicolumn{1}{c}{\bf Name} & {\bf RA} &{\bf Dec} & {\bf Sep.} & {\bf Sp} & {\bf Vp} & {\bf Vr} & {\bf Int} & {\bf notes}\\
\multicolumn{1}{c}{\bf ($l,b$)} & {\bf (J2000)}  & {\bf (J2000)}&  &  \multicolumn{1}{c}{\bf and} & \multicolumn{1}{c}{\bf ($l,b$)} & {\bf (J2000)}  & {\bf (J2000)} & {\bf ('')} &{\bf (Jy)}&   &     \\	
\multicolumn{1}{c}{\bf (degrees)}& {\bf (h m s)}&{\bf ($^{o}$ $'$ $``$)}&  &\multicolumn{1}{c}{\bf epoch} &  \multicolumn{1}{c}{\bf (degrees)}& {\bf (h m s)}&{\bf ($^{o}$ $'$ $``$)} \\ \hline \hline\\

G\,213.61--12.6			&06:07:47.9	&--06:22:57		&3.7E--4	&H214dec07	& G\,213.705--12.597	&	06 07 47.85	&	--06 22 56.6	&	1	&	92	& 12	&	4,13	&	78 & Mon R2\\
G\,269.45--1.47			&09:03:13.5	&--48:55:22		&0.15	& H214jul08		&G\,269.461--1.471	&	09 03 15.01   & --48 55 31.4	&	18 	& 0.6		& 51	& 49,52 & 1.6	\\%269.45$\_$2\\
						&			&				&		                & H214jul08    & 	 G\,269.457--1.467		&	09 03 14.87	& 	--48 55 12.6   	&	 24	&	0.8	& 66	&	56,69	&	1.7\\
\end{tabular}\label{tab:dust_sources} 
\end{center}
\end{table}

\clearpage

\begin{table}\addtocounter{table}{-1}
\begin{center}
  \caption{-- {\emph {continued}}}

\begin{tabular}{lccrllcccllcrl}\hline
    \multicolumn{3}{c}{\bf Dust clump} & &  \multicolumn{10}{c}{\bf Water maser} \\
 \multicolumn{1}{c}{\bf Name} & {\bf RA}  & {\bf Dec} & \multicolumn{1}{c}{\bf prob.} & \multicolumn{1}{c}{\bf array} &  \multicolumn{1}{c}{\bf Name} & {\bf RA} &{\bf Dec} & {\bf Sep.} & {\bf Sp} & {\bf Vp} & {\bf Vr} & {\bf Int} & {\bf notes}\\
\multicolumn{1}{c}{\bf ($l,b$)} & {\bf (J2000)}  & {\bf (J2000)}&  &  \multicolumn{1}{c}{\bf and} & \multicolumn{1}{c}{\bf ($l,b$)} & {\bf (J2000)}  & {\bf (J2000)} & {\bf ('')} &{\bf (Jy)}&   &     \\	
\multicolumn{1}{c}{\bf (degrees)}& {\bf (h m s)}&{\bf ($^{o}$ $'$ $``$)}&  &\multicolumn{1}{c}{\bf epoch} &  \multicolumn{1}{c}{\bf (degrees)}& {\bf (h m s)}&{\bf ($^{o}$ $'$ $``$)} \\ \hline \hline\\
G\,291.630--0.545			&11:15:08.9	&--61:17:08		&0.89 		&	BCEP03	&	G\,291.629--0.541	&	11 15 08.88	&	--61 16 54.8	&	13	&	70	&11	&	8,16	\\
G\,291.614--0.443			&11:15:19.9	&--61:11:08		&5.9E--5		& 6Baug08	&	G\,291.618--0.442	&	11 15 21.82	&	--61 11 09.3 &		14	&	0.7	&	15	& 0,15	&	2.4	& slightly spread\\%	 291.614-0.44$\_2$ \\
%11 15 21.33	& --61 11 19.1	& 291.614-0.44$\_1$	& too weak to extract a decent spectrum - probably ditch\\
%& 	&	&	&	&		11 15 21.88	&	--61 11 08.8 & 291.614-0.44$\_3$ & just the one feature, also emission at this vel from previous source\\
& 	&	&	&	6Baug08	&	G\,291.614--0.446	&	11 15 19.45	&	--61 11 15.2 &	 8	&	0.3	&	7	&	7,8	&	0.2\\
\end{tabular}
\end{center}
\end{table}

\clearpage

\begin{table}\addtocounter{table}{-1}
\begin{center}
  \caption{-- {\emph {continued}}}
%   \vspace{0.5cm}
    \footnotesize
\begin{tabular}{lccrllcccllcrl}\hline
    \multicolumn{3}{c}{\bf Dust clump} & &  \multicolumn{10}{c}{\bf Water maser} \\
 \multicolumn{1}{c}{\bf Name} & {\bf RA}  & {\bf Dec} & \multicolumn{1}{c}{\bf prob.} & \multicolumn{1}{c}{\bf array} &  \multicolumn{1}{c}{\bf Name} & {\bf RA} &{\bf Dec} & {\bf Sep.} & {\bf Sp} & {\bf Vp} & {\bf Vr} & {\bf Int} & {\bf notes}\\
\multicolumn{1}{c}{\bf ($l,b$)} & {\bf (J2000)}  & {\bf (J2000)}&  &  \multicolumn{1}{c}{\bf and} & \multicolumn{1}{c}{\bf ($l,b$)} & {\bf (J2000)}  & {\bf (J2000)} & {\bf ('')} &{\bf (Jy)}&   &     \\	
\multicolumn{1}{c}{\bf (degrees)}& {\bf (h m s)}&{\bf ($^{o}$ $'$ $``$)}&  &\multicolumn{1}{c}{\bf epoch} &  \multicolumn{1}{c}{\bf (degrees)}& {\bf (h m s)}&{\bf ($^{o}$ $'$ $``$)} \\ \hline \hline\\
G\,305.520--0.020			&13:14:06.1	&--62:46:41		&5.9E--5	& 6Baug08	&  & & & & $<$135& & --210,140\\
G\,305.549+0.002			&13:14:20.0	&--62:45:13		&2.8E--4& 6Baug08	&  & & & &$<$112     & & --210,140\\
G\,305.552+0.013			&13:14:21.2	&--62:44:33		&1.6E--4	& 6Baug08	&  & & & & $<$162 & & --210,140\\
\end{tabular}
\end{center}
\end{table}

\clearpage

\begin{table}\addtocounter{table}{-1}
\begin{center}
  \caption{-- {\emph {continued}}}
%   \vspace{0.5cm}
    \footnotesize
\begin{tabular}{lccrllcccllcrl}\hline
    \multicolumn{3}{c}{\bf Dust clump} & &  \multicolumn{10}{c}{\bf Water maser} \\
 \multicolumn{1}{c}{\bf Name} & {\bf RA}  & {\bf Dec} & \multicolumn{1}{c}{\bf prob.} & \multicolumn{1}{c}{\bf array} &  \multicolumn{1}{c}{\bf Name} & {\bf RA} &{\bf Dec} & {\bf Sep.} & {\bf Sp} & {\bf Vp} & {\bf Vr} & {\bf Int} & {\bf notes}\\
\multicolumn{1}{c}{\bf ($l,b$)} & {\bf (J2000)}  & {\bf (J2000)}&  &  \multicolumn{1}{c}{\bf and} & \multicolumn{1}{c}{\bf ($l,b$)} & {\bf (J2000)}  & {\bf (J2000)} & {\bf ('')} &{\bf (Jy)}&   &     \\	
\multicolumn{1}{c}{\bf (degrees)}& {\bf (h m s)}&{\bf ($^{o}$ $'$ $``$)}&  &\multicolumn{1}{c}{\bf epoch} &  \multicolumn{1}{c}{\bf (degrees)}& {\bf (h m s)}&{\bf ($^{o}$ $'$ $``$)} \\ \hline \hline\\
G\,0.325--0.242			&17:47:20.1	&--28:47:04		&7.8E--5 		& 6Baug08	& & & & &	$<$115	&	& --150,200\\%0.324-0.241		\\
G\,1.124--0.065			&17:48:31.6	&--28:00:31		&2.2E--3		& 6Baug08	& & & & & 	$<$240	& 	& --150,200\\%1.129-0.068\\
G\,1.134--0.073			&17:48:34.7	&--28:00:16		&3.1E--4		&6Baug08	& & & & &	$<$240	& 	& --150,200\\
\end{tabular}
\end{center}
\end{table}

\clearpage

\begin{table}\addtocounter{table}{-1}
\begin{center}
  \caption{-- {\emph {continued}}}
%   \vspace{0.5cm}
    \footnotesize
\begin{tabular}{lccrllcccllcrl}\hline
    \multicolumn{3}{c}{\bf Dust clump} & &  \multicolumn{10}{c}{\bf Water maser} \\
 \multicolumn{1}{c}{\bf Name} & {\bf RA}  & {\bf Dec} & \multicolumn{1}{c}{\bf prob.} & \multicolumn{1}{c}{\bf array} &  \multicolumn{1}{c}{\bf Name} & {\bf RA} &{\bf Dec} & {\bf Sep.} & {\bf Sp} & {\bf Vp} & {\bf Vr} & {\bf Int} & {\bf notes}\\
\multicolumn{1}{c}{\bf ($l,b$)} & {\bf (J2000)}  & {\bf (J2000)}&  &  \multicolumn{1}{c}{\bf and} & \multicolumn{1}{c}{\bf ($l,b$)} & {\bf (J2000)}  & {\bf (J2000)} & {\bf ('')} &{\bf (Jy)}&   &     \\	
\multicolumn{1}{c}{\bf (degrees)}& {\bf (h m s)}&{\bf ($^{o}$ $'$ $``$)}&  &\multicolumn{1}{c}{\bf epoch} &  \multicolumn{1}{c}{\bf (degrees)}& {\bf (h m s)}&{\bf ($^{o}$ $'$ $``$)} \\ \hline \hline\\
G\,8.718--0.410			&18:06:37.3	&--21:36:33		&2.4E--5 &	6Baug08	&	& & & & $<$113	&	& --140,210\\%8.718-0.408\\
G\,9.966--0.020			&18:07:45.8	&--20:19:47		&3.1E--5 	&	6Baug08	&	& & & & $<$128	&	& --140,210\\%9.965-0.020	\\
G\,9.99--0.03				&18:07:50.4	&--20:18:51		&1.5E--3	&	6Baug08	&	G\,9.986--0.028	& 18 07 50.12	&	--20 18 56.6	&7	& 14	& 46 & 40,60 & 38\\
\end{tabular}
\end{center}
\end{table}

\clearpage

\begin{table}\addtocounter{table}{-1}
\begin{center}
  \caption{-- {\emph {continued}}}
 %  \vspace{0.5cm}
    \footnotesize
\begin{tabular}{lccrllcccllcrl}\hline
    \multicolumn{3}{c}{\bf Dust clump} & &  \multicolumn{10}{c}{\bf Water maser} \\
 \multicolumn{1}{c}{\bf Name} & {\bf RA}  & {\bf Dec} & \multicolumn{1}{c}{\bf prob.} & \multicolumn{1}{c}{\bf array} &  \multicolumn{1}{c}{\bf Name} & {\bf RA} &{\bf Dec} & {\bf Sep.} & {\bf Sp} & {\bf Vp} & {\bf Vr} & {\bf Int} & {\bf notes}\\
\multicolumn{1}{c}{\bf ($l,b$)} & {\bf (J2000)}  & {\bf (J2000)}&  &  \multicolumn{1}{c}{\bf and} & \multicolumn{1}{c}{\bf ($l,b$)} & {\bf (J2000)}  & {\bf (J2000)} & {\bf ('')} &{\bf (Jy)}&   &     \\	
\multicolumn{1}{c}{\bf (degrees)}& {\bf (h m s)}&{\bf ($^{o}$ $'$ $``$)}&  &\multicolumn{1}{c}{\bf epoch} &  \multicolumn{1}{c}{\bf (degrees)}& {\bf (h m s)}&{\bf ($^{o}$ $'$ $``$)} \\ \hline \hline\\
G\,12.88+0.48			&18:11:51.4	&--17:31:30		&0.025	&	6Baug08	&	G\,12.889+0.489	& 18 11 51.46	&	--17 31 28.9	& 1 & 8 & 29 & 28,32 & 16\\%	12.915+0.491$\_$2	\\
G\,11.948--0.003			&18:11:52.9	&--18:36:03		&4.8E--3		&	6Baug08	&	& & & & $<$109	& & --140,210\\%	11.948-0.033				\\
G\,12.914+0.493			&18:11:53.6	&--17:30:02		&1.4E--4	&	6Baug08	&	G\,12.915+0.493	& 18 11 53.74	&	--17 29 58.8	& 4 & 2.7 & 33 & 31,34 & 3.9 \\
\end{tabular}
\end{center}
\end{table}

\clearpage

\begin{table}\addtocounter{table}{-1}
\begin{center}
  \caption{-- {\emph {continued}}}
 %  \vspace{0.5cm}
    \footnotesize
\begin{tabular}{lccrllcccllcrl}\hline
    \multicolumn{3}{c}{\bf Dust clump} & &  \multicolumn{10}{c}{\bf Water maser} \\
 \multicolumn{1}{c}{\bf Name} & {\bf RA}  & {\bf Dec} & \multicolumn{1}{c}{\bf prob.} & \multicolumn{1}{c}{\bf array} &  \multicolumn{1}{c}{\bf Name} & {\bf RA} &{\bf Dec} & {\bf Sep.} & {\bf Sp} & {\bf Vp} & {\bf Vr} & {\bf Int} & {\bf notes}\\
\multicolumn{1}{c}{\bf ($l,b$)} & {\bf (J2000)}  & {\bf (J2000)}&  &  \multicolumn{1}{c}{\bf and} & \multicolumn{1}{c}{\bf ($l,b$)} & {\bf (J2000)}  & {\bf (J2000)} & {\bf ('')} &{\bf (Jy)}&   &     \\	
\multicolumn{1}{c}{\bf (degrees)}& {\bf (h m s)}&{\bf ($^{o}$ $'$ $``$)}&  &\multicolumn{1}{c}{\bf epoch} &  \multicolumn{1}{c}{\bf (degrees)}& {\bf (h m s)}&{\bf ($^{o}$ $'$ $``$)} \\ \hline \hline\\
G\,15.027--0.651			&18:20:18.1	&--16:11:15    		&1.6E--4&	6Baug08	&	G\,14.892--0.730	& 18 20 19.60	&	--16 10 38	.4	& 43	& 4.3 & 25 & 20,33 & 14 &\\%	15.037-0.626$\_$1	& this *might be a s/l?'\\
G\,15.054--0.641			&18:20:19.2	&--16:09:31		&2.6E--5	&	6Baug08	&	& & & & $<$358	& & --140,210\\%	15.037-0.626	   \\ 
G\,14.983--0.687			&18:20:20.9	&--16:14:35		&2.1E--4	&	6Baug08	&	& & & & $<$199	& & --140,210\\
\end{tabular}
\end{center}
\end{table}

\clearpage

\begin{table}\addtocounter{table}{-1}
\begin{center}
  \caption{-- {\emph {continued}}}
 %  \vspace{0.5cm}
    \footnotesize
\begin{tabular}{lccrllcccllcrl}\hline
    \multicolumn{3}{c}{\bf Dust clump} & &  \multicolumn{10}{c}{\bf Water maser} \\
 \multicolumn{1}{c}{\bf Name} & {\bf RA}  & {\bf Dec} & \multicolumn{1}{c}{\bf prob.} & \multicolumn{1}{c}{\bf array} &  \multicolumn{1}{c}{\bf Name} & {\bf RA} &{\bf Dec} & {\bf Sep.} & {\bf Sp} & {\bf Vp} & {\bf Vr} & {\bf Int} & {\bf notes}\\
\multicolumn{1}{c}{\bf ($l,b$)} & {\bf (J2000)}  & {\bf (J2000)}&  &  \multicolumn{1}{c}{\bf and} & \multicolumn{1}{c}{\bf ($l,b$)} & {\bf (J2000)}  & {\bf (J2000)} & {\bf ('')} &{\bf (Jy)}&   &     \\	
\multicolumn{1}{c}{\bf (degrees)}& {\bf (h m s)}&{\bf ($^{o}$ $'$ $``$)}&  &\multicolumn{1}{c}{\bf epoch} &  \multicolumn{1}{c}{\bf (degrees)}& {\bf (h m s)}&{\bf ($^{o}$ $'$ $``$)} \\ \hline \hline\\
G\,23.420--0.235			&18:34:48.4	&--08:33:57		&3.1E--4	& H214jul08	& & & & & $<$5000	& 	  & --120,230\\%	       	  & 23.395-0.195	      & not in linmos image so very offset from feild centre!\\
G\,23.319--0.298			&18:34:50.7	&--08:41:03		&4.1E--4	& H214jul08	& G\,23.320--0.298	&	18 34 50.62 & --08 41 00.1 & 3	&	0.2 & 99 & 98,99 & 0.1\\%		23.319-0.298$\_$1\\
G\,23.754+0.095			&18:36:06.1	&--07:13:47		&2.5E--3	& H214jul08	& & & & & $<$147	& & --120,230\\
\hline
\end{tabular}
\end{center}
\end{table}
\end{landscape}

\begin{figure*}
	\psfig{figure=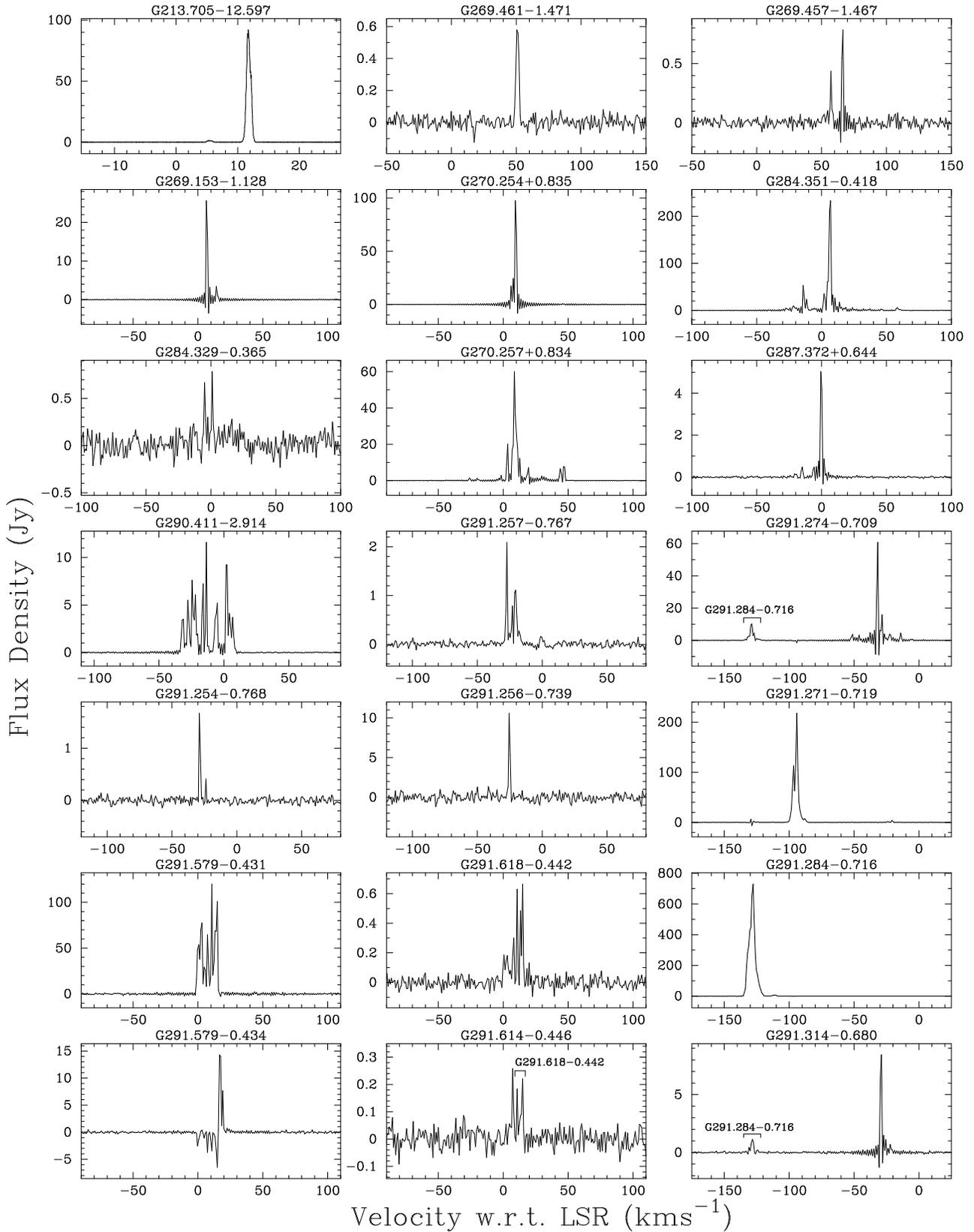,width=17cm}
\caption{Spectra of the 22-GHz water masers detected towards 1.2-mm dust clumps.}
\label{fig:dust_water_spectra}
\end{figure*}

\begin{figure*}\addtocounter{figure}{-1}
	\psfig{figure=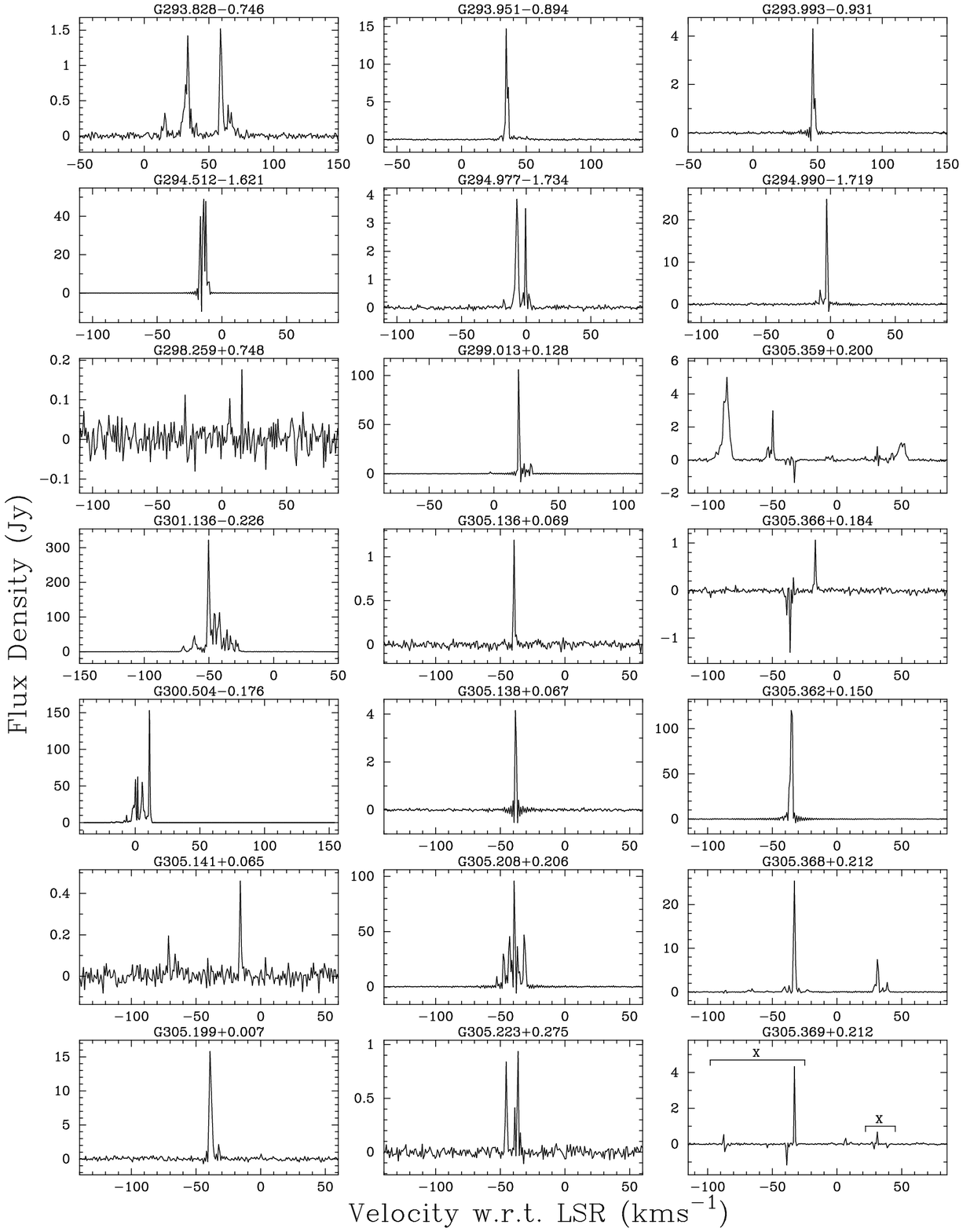,width=17cm}
\caption{--{\emph {continuued}}}
\end{figure*}

\begin{figure*}\addtocounter{figure}{-1}
	\psfig{figure=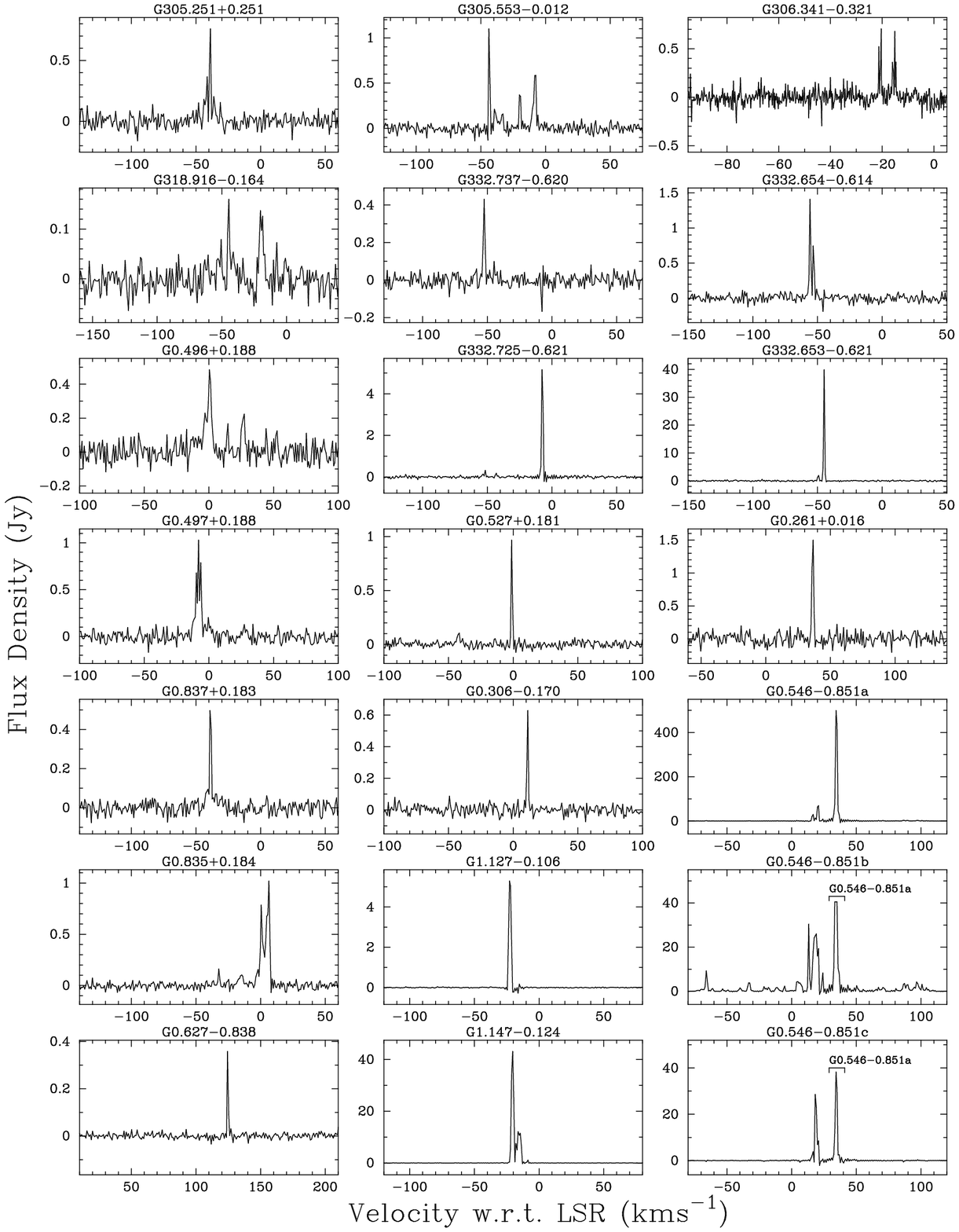,width=17cm}
\caption{--{\emph {continuued}}}
\end{figure*}

\begin{figure*}\addtocounter{figure}{-1}
	\psfig{figure=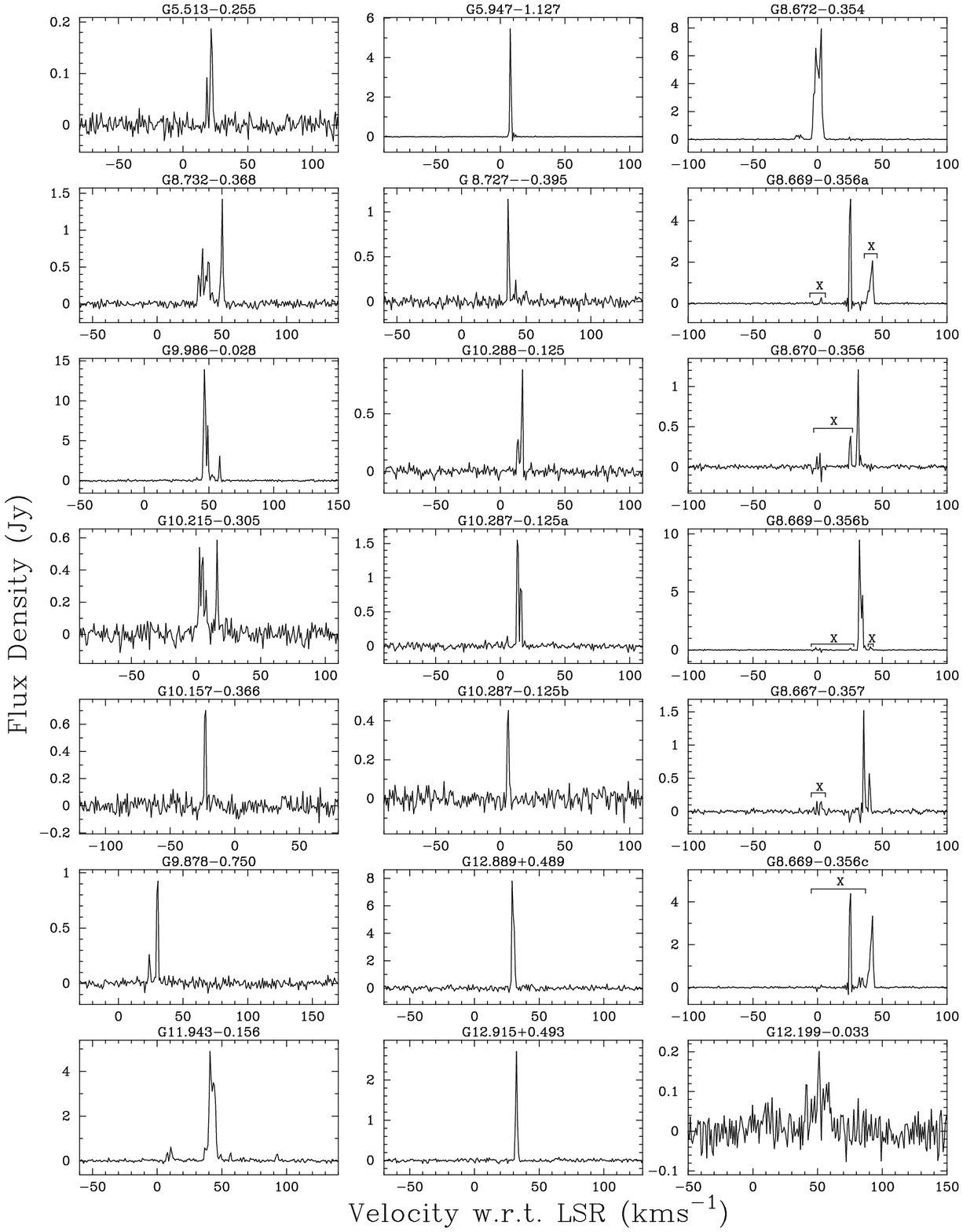,width=17cm}
\caption{--{\emph {continuued}}}
\end{figure*}

\begin{figure*}\addtocounter{figure}{-1}
	\psfig{figure=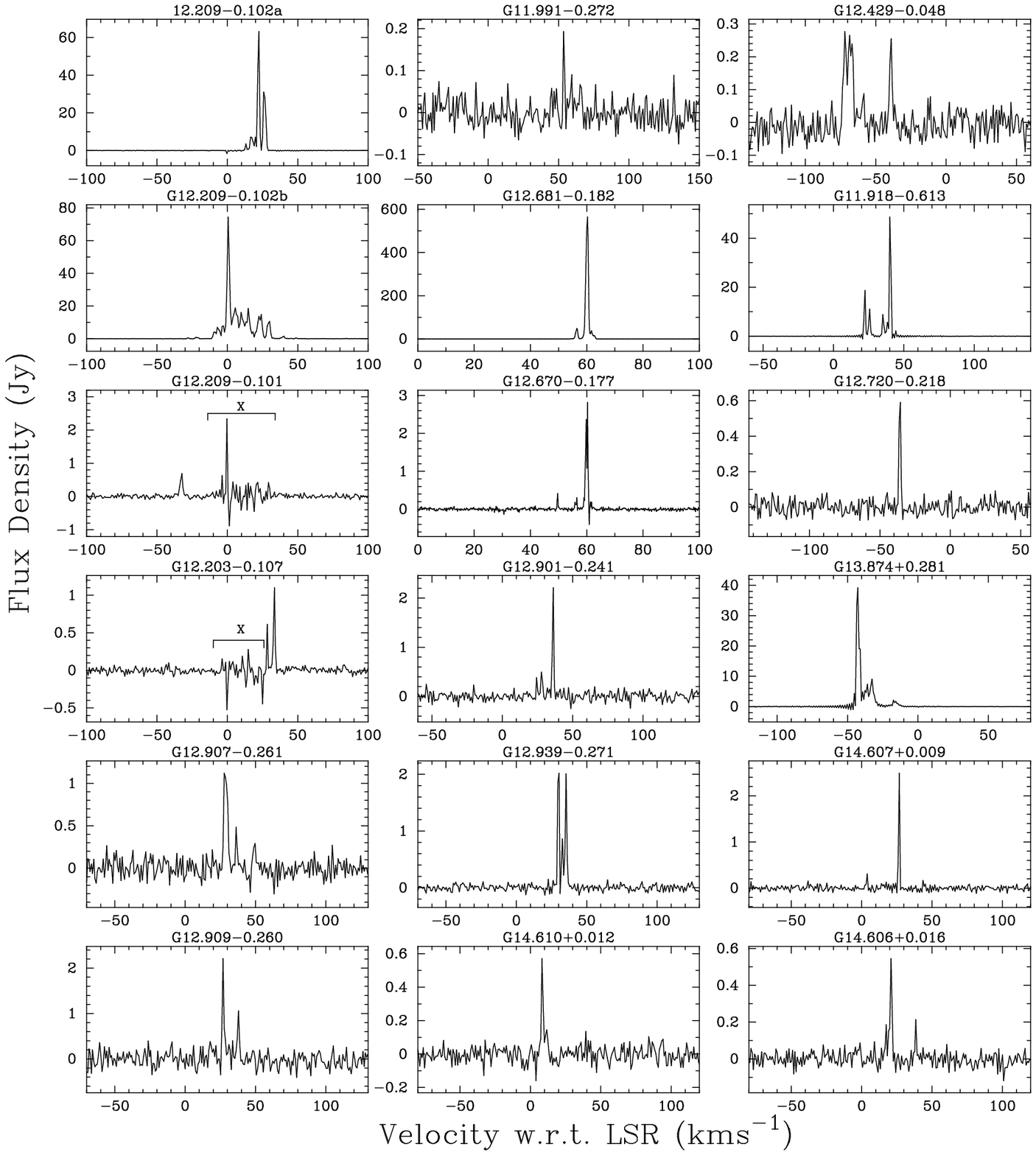,width=17cm}
\caption{--{\emph {continuued}}}
\end{figure*}

\begin{figure*}\addtocounter{figure}{-1}
	\psfig{figure=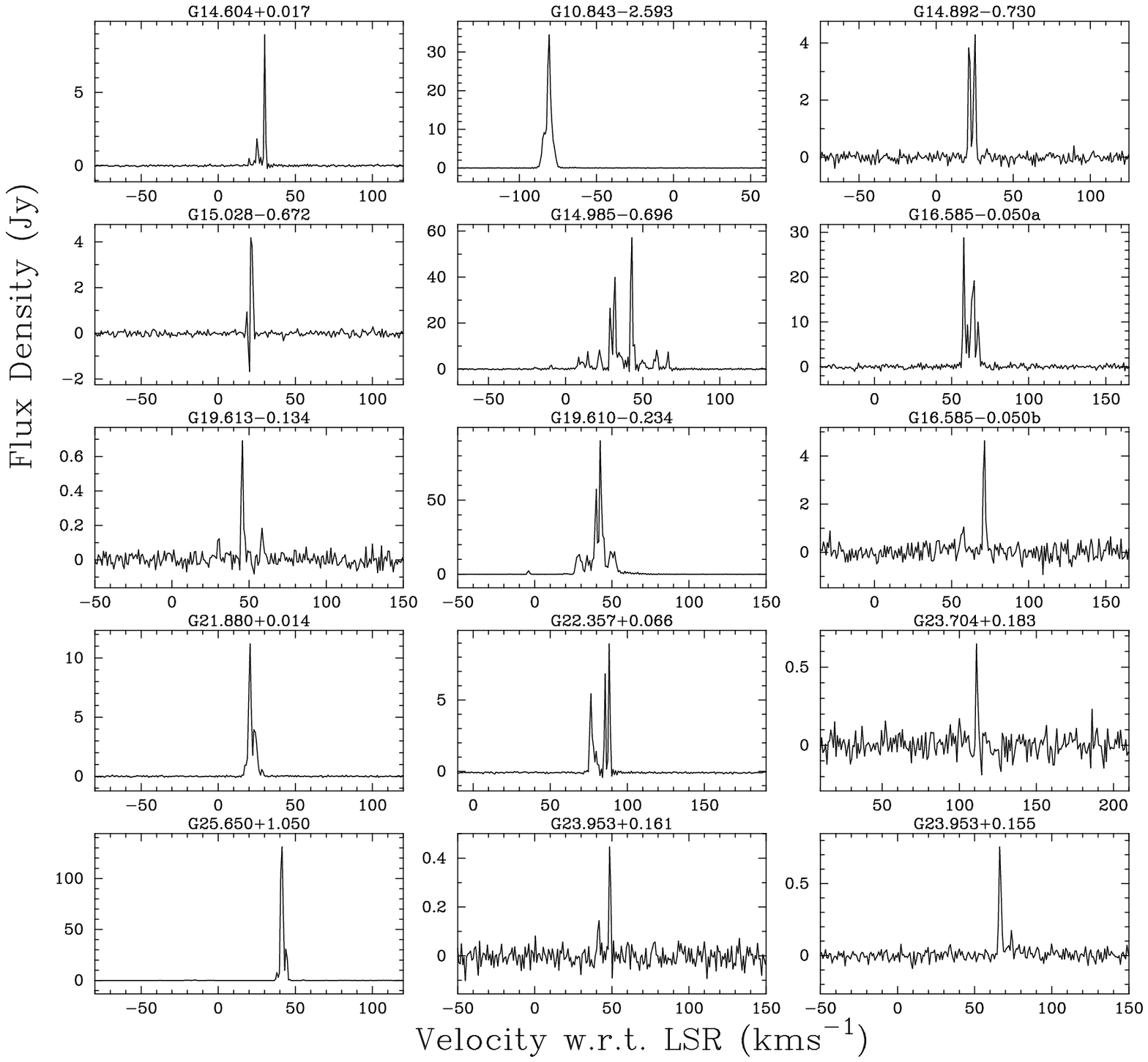,width=17cm}
\caption{--{\emph {continuued}}}
\end{figure*}
	
\clearpage
		
\begin{figure*}\addtocounter{figure}{-1}
	\psfig{figure=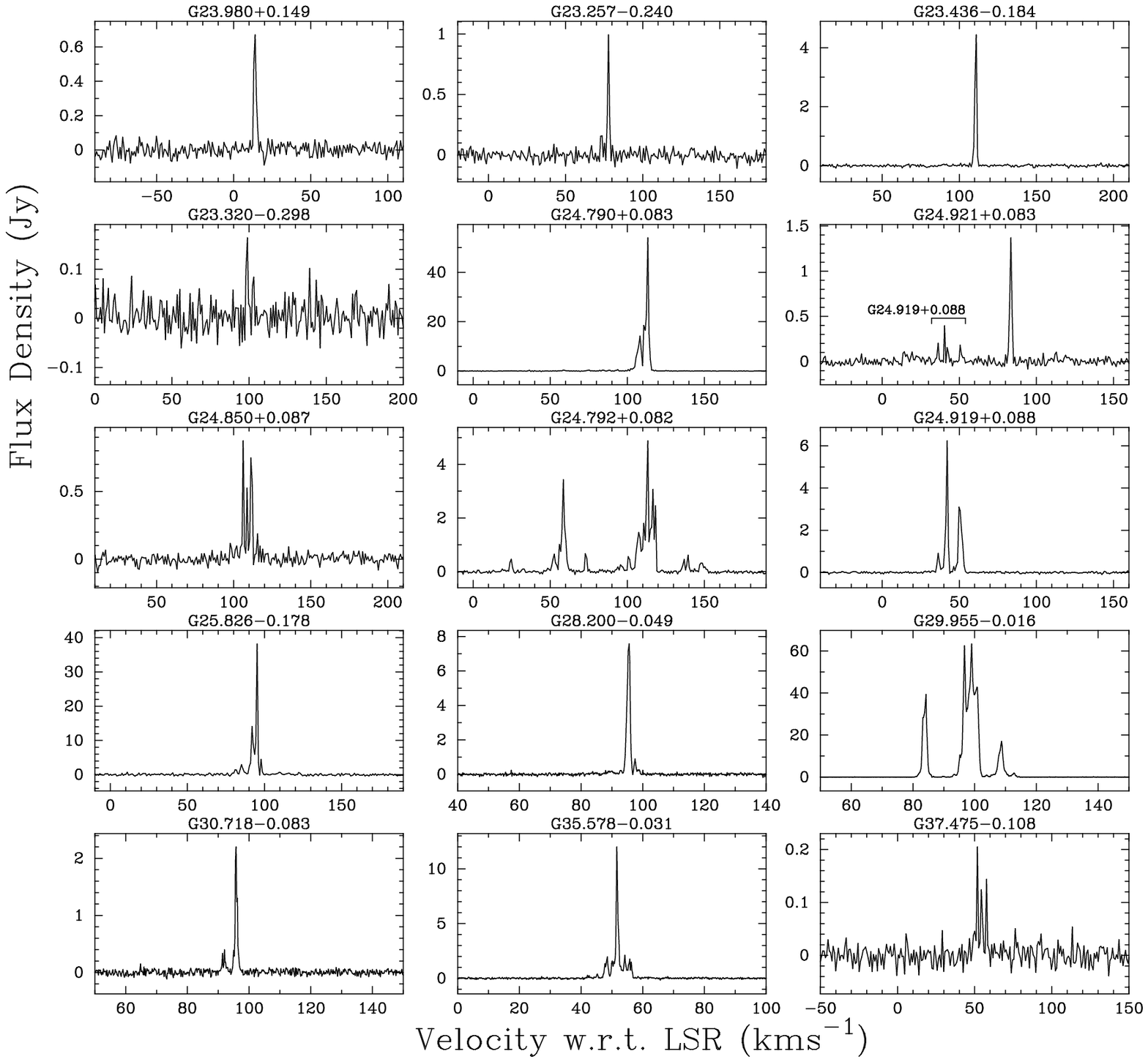,width=17cm}
\caption{--{\emph {continuued}}}
\end{figure*}

\subsection{Individual sources}\label{sect:water_dust_ind}

Here we draw attention to sources that we were unable to adequately describe in Table~\ref{tab:dust_sources}. Close neighbouring water masers are discussed, as well as sources with surprising velocities and intriguing histories. For some sources, associations with other maser species are also discussed.

\subparagraph{G\,213.705--12.597.} This water maser is located in a nearby star formation region, Mon R2.  Associated with this star formation region are many maser species, including 6.7- and 12.2-GHz methanol, main-line OH, and some excited OH maser transitions. Of particular interest is the excited OH maser at the 4765-MHz transition; these sources are very rare (only about 20 known in the Galaxy) and this source is by far the strongest known at this transition. Furthermore, the emission from this 4765-MHz OH maser undergoes flaring activity that
has not been seen in any other maser of this type \citep{Smits03,Smits98}. %distance

\subparagraph{G\,284.351--0.418.} This source is also presented in \citet{Breen10b} as it is coincident with an OH maser \citep{C98}, although is named G\,284.350--0.418, due to a slight difference in measured position affecting the rounded Galactic coordinates. In \citet{Breen10b}, spectra from two epochs are shown to display the typical level of variability seen in the water maser sources. When the 2003 and 2004 spectra are compared to our current observations, it can be seen that the source has once again varied by a moderate level and has increased in peak flux density by about a factor of 2.

\subparagraph{G\,291.271--0.719, G\,291.274--0.709 and G\,291.284--0.716.} These three sources are also listed in \citet{Breen10b} since they all fall within the field of a target OH maser source. \citet{CP08} also observed this group of sources, and remark that the strongest source, G\,291.284--0.716, is a member of a distinct class of sources that are dominated by strong blue shifted outflows. In fact, maser emission has never been detected at the systemic velocity of the source (to a limit of 0.3~Jy in these observations), which is around 100 \kms from the detected emission. The location of G\,291.284--0.716 is very close to the edge of the 1.2-mm dust clump emission, unlike the majority of sources.

\subparagraph{G\,291.579--0.431 and G\,291.579--0.434.} G\,291.579--0.431 shows emission spread over approximately 2 arcsec and in Table~\ref{tab:dust_sources} we present the median of the positions measured for the individual features. This source is clearly separated from G\,291.579--0.434, which is located more than 10 arcsec away.

\subparagraph{G\,301.136--0.226.} As noted in \citet{Breen10b}, emission from this source is significantly spread. In the observations carried out in 2008 July, we observe individual maser components spread over almost 5 arcsec. An OH maser is located within the spread of the water maser emission \citep{C98}.

\subparagraph{G\,305.362+0.150.} During the observations in 2008 August we detected a moderately strong water maser exhibiting one main spectral feature. In 2004 (presented in \citet{Breen10b}) an additional, broader feature of more than 100~Jy was also detected and has disappeared intervening interval. 

\subparagraph{G\,0.546--0.851b.} This water maser falls within a cluster of three distinct water maser sites and has a velocity range that exceeds 170 \kmsns. The emission extends almost symmetrically about the peak water maser emission. 

\subparagraph{G\,5.513--0.255.} Like G\,291.284--0.716, this water masers lies unusually close to the edge of the 1.2-mm dust clump emission that was observed by \citet{Hill05}. No other similarities between the sources are evident. This source is very weak and shows emission close to the likely systemic velocity of the region, while G\,291.284--0.716 is very strong and is a dominant blue-shifted source.

\subparagraph{G\,8.727--0.395.} This water maser is clearly projected against 1.2-mm dust continuum emission on inspection of the corresponding dust map of \citet{Hill05}, although no dust clump is reported at this position. It is likely that this source was omitted from the catalogue of \citet{Hill05} in error.

\subparagraph{G\,10.473+0.027 and G\,10.480+0.034.} The observations presented in Table~\ref{tab:dust_sources} have been taken from \citet{Breen10b}. Both of the sources were detected in observations carried out in 2003, but G\,10.480+0.034 had fallen below the detection limits in 2004 ($<$0.2~Jy).

\subparagraph{G\,11.903--0.142.} The observations presented in Table~\ref{tab:dust_sources} have been taken from the observations carried out in \citet{Breen10b}, and similarly to G\,10.480+0.034, this water maser was detected in 2003, but by 2004 the emission had fallen below our detection limits. Since the emission detected in 2003 had a peak flux density of 0.3~Jy, this is not surprising.

\subparagraph{G\,12.209--0.101, G\,12.429--0.048, G\,12.720--0.218 and G\,13.874+0.281.} These four well separated water maser sources all exhibit water maser emission at only negative velocities. From the velocities of nearby methanol masers, systemic velocities in this region of the sky would be expected to be between 20 - 60 \kmsns. Therefore these sources are all potential candidates for the class of sources that are dominated by blue-shifted emission.

\subsection{Water masers with no apparent associated dust continuum emission}\label{sect:special}

In addition to the water maser sources presented in Table~\ref{tab:dust_sources} and Fig.~\ref{fig:dust_water_spectra}, we detect several other sources which are of special interest. While an exhaustive attempt to search for water maser emission beyond the boundaries of the 1.2-mm dust clumps was not made, here we list the additional water maser sources that were identified.

\begin{figure*}
	\psfig{figure=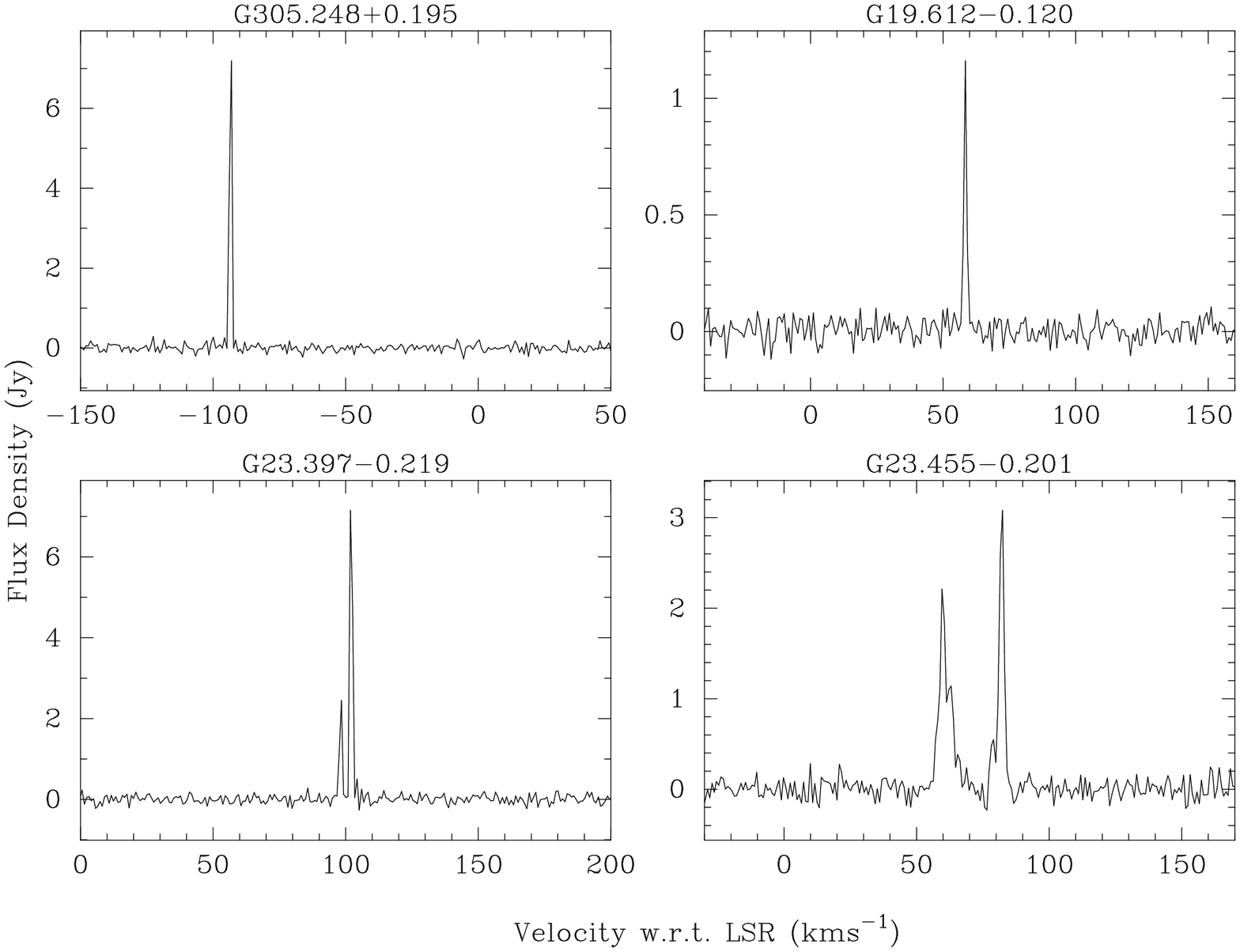,width=11cm,angle=270}
\caption{Water maser sources with no associated 1.2-mm dust clump emission, detected within the target fields.}
\label{fig:special_nodust}
\end{figure*}

Interestingly, we find four water maser sources that appear not to be associated with any 1.2-mm dust clump emission and these are presented in Table~\ref{tab:no_dust} and Fig.~\ref{fig:special_nodust}. \citet{Hill05} detected 1.2-mm dust clump emission in the fields of all of the target 6.7-GHz methanol masers and \UCHII regions, but for 20 of the methanol masers and 9 of the \UCHII regions there is no spatially coincident dust continuum emission. Investigation by \citet{Hill05} showed that these sources appeared to be no different to those where 1.2-mm dust continuum was detected. Since the presence of 1.2-mm dust clump emission indicates the presence of cold, deeply embedded sources, two of the possible explanations put forward by \citet{Hill05} are that these sources are more evolved and hence no longer in the cold core phase, or, perhaps they are less massive.

We have investigated possible mid-infrared counterparts for these four sources by comparing their locations with products from the GLIMPSE survey. All are projected against regions of extended infrared emission, and two sources (G\,19.612--0.120 and G\,23.455--0.201) are coincident to within 1 arcsec of a GLIMPSE point source. Interestingly, the two sources with associated point sources are also associated with methanol masers. This suggests that these two sources may be located at the far distance and consequently were not detected in 1.2-mm dust continuum due to sensitivity limitations. The nature of the other two sources is more confusing and certainly warrants future investigation.

Unlike 6.7-GHz methanol masers, water masers have been detected towards both evolved and low-mass stars. Their presence in locations devoid of 1.2-mm dust continuum emission is therefore much more easily accounted for than methanol masers. Comments on each of these four water masers are given below.

\subparagraph{G\,305.248+0.195.} This water maser is separated from the boundary of the nearest dust clump by more than an arcminute.

This source falls beyond the latitude range completely searched for the presence of water maser emission by \citet{CB10} which was focused on a region bounded by a longitude range of 305.0 - 306.26 $^{\circ}$ and $\pm$ 0.15$^{\circ}$ latitude. A total of 23 water maser sources were detected in this portion of the Galactic plane. Analysis of the masers found in this area, along with another region (longitude range of 311.0 - 312.18$^{\circ}$ and $\pm$ 0.15$^{\circ}$ latitude) showed a surprising preponderance of sources with highly blue-shifted emission. Investigation by \citet{CB10} led to a suggestion that these water masers represent a population of sources that are tracing a very early evolutionary stage, perhaps preceding the stage where 6.7-GHz methanol masers are seen.

Interesting, G\,305.248+0.195 only shows emission at a blue-shifted velocity of --95 \kms (nearby methanol masers detected in the Methanol Multibeam Survey exhibit peak velocities between --40 and --30 \kmsns), and therefore adds to the already high number of blue-shifted sources in this region of the Galactic plane. However, the interpretation that such sources may be very young directly contradicts one of the suggestions given by \citet{Hill05} as to why some sources are devoid of 1.2-mm dust continuum emission - that they are more evolved. Alternatively, it is possible that this water maser is associated with a low-mass object, although a high velocity feature $>$ 50 \kms from the systemic velocity of the region would be unusual for a low-mass star. 

\subparagraph{G\,19.612--0.120.} This water maser is coincident with one of the methanol masers from \citet{Walsh98} that was targeted by \citet{Hill05}, but is also offset from the boundary of the nearest dust clump by about an arcminute. \citet{GreenMMB10} present evidence suggesting that this methanol maser is located at the far distance, which would limit the sensitivity of the 1.2-mm dust continuum observations, but not to the extent that no detection of dust continuum emission would be expected.

\subparagraph{G\,23.397--0.219 and G\,23.455--0.201.} Both of these water masers fall within the one dust continuum emission map, but neither are directly associated with any detectable dust continuum emission. The first source is apparently solitary (i.e. without any associated \UCHII region or 6.7-GHz methanol maser emission), but the second source appears to be coincident with an \UCHII region reported by \citet{Walsh98}.

\begin{table*}
\begin{center}
  \caption[Four water maser sources detected in the target fields that are not coincident with any detected 1.2-mm dust clump emission.]{Four water maser sources detected in the target fields that are not coincident with any detected 1.2-mm dust clump emission. Columns 1-7 give the: Water maser source name; water maser right ascension; declination; peak flux density (Jy); velocity of the water maser peak emission (\kmsns); velocity range (\kmsns); and the integrated flux density of the water maser emission (Jy \kmsns), respectively.}
 %  \vspace{0.5cm}
\begin{tabular}{lccrcccccllcrl}\hline
\multicolumn{1}{c}{\bf Name} & {\bf RA} &{\bf Dec}  & {\bf Sp} & {\bf Vp} & {\bf Vr}& {\bf Int} \\
\multicolumn{1}{c}{\bf ($l,b$)} & {\bf (J2000)}  & {\bf (J2000)}  &{\bf (Jy)}& {\bf (\kmsns)}  & {\bf (\kmsns)} & {\bf (Jy \kmsns)}   \\	
\multicolumn{1}{c}{\bf (degrees)}& {\bf (h m s)}&{\bf ($^{o}$ $'$ $``$)} \\ \hline \hline
G\,305.248+0.195	&13 11 34.84 & --62 35 11.0	& 7 & --93 & --95,--92 & 9 \\%& BS outflow, no dust\\
G\,19.612--0.120	& 18 27 13.55 & --11 53 14.7 & 1.2	&	58 & 57,60 & 1.6 \\%& no dust but coincident with meth?\\ %19.609-0.135$\_$2 
G\,23.397--0.219	& 18 34 42.32 & --08 34 43.3	& 7 	&	102 & 97,106 & 14 \\%& no dust\\ %23.395-0.195_1.spt
G\,23.455--0.201	& 18 34 44.89 & --08 31 07.4	& 3.1& 	82 & 57,84 & 17 \\  \hline %& no dust but associated with a UCHii region\\ %23.438-0.184$\_$2		
%*** Also note 2 nearby single-feature sources located at 18 14 20.264 --17 54 13.3 (12.853-0.224$\_$1) and 18 14 20.18 --17 54 13.8 (12.853-0.224$\_$2) they were found in a cube towards a source with no radius measurement -> no entry in the table...	
% extra& 	&	&	&	6Baug08&	13 11 42.19	& --62 49 17.5 & 305.233-0.02$\_$1 & off the edge of the dust clump map\\ 
	
	\end{tabular}\label{tab:no_dust}
\end{center}
\end{table*}

\section{Discussion}		

\subsection{Detection statistics}

Altogether, we find water maser emission towards 128 of the 294 1.2-mm dust clumps, equating to a detection rate of 44 per cent. Investigation of the detection rate of water masers towards 1.2-mm dust clumps shows some variation with Galactic longitude. The water maser detection rate towards dust clumps located between longitudes of 213 and 333 degrees is 55 per cent (51 of 93), compared with 38 per cent (77 of 201) for the longitude range 0 to 38 degrees. Although not covering exactly the same longitude ranges, it is curious to note that the detection rates of 12.2-GHz methanol masers in the longitude range 270 to 305 degrees were significantly lower than in other parts of the Galaxy \citep{Breen12stats}. If it were the relative distances that were affecting our detection rates then, if anything, we would expect to have a lower detection rate in the 213 to 333 degree longitude region as we would expect relatively more dust clump detections in these regions. Since our maser observations are of a high sensitivity we would expect a minimal detection rate dependence on distance. \citet{CB10} and \citet{CBE10} completed systematic searches for water masers within three regions of the Galactic plane. They show that there is a large difference in the water maser population densities within these regions, suggesting that local factors have an effect on the detection statistics. It is therefore possible that there are broader localised effects contributing to our detection statistics here. This warrants further investigation, but is beyond the scope of this paper.

%## cut 25 Jan
%Figure~\ref{fig:long_lat_dust} shows the distribution of the targeted 1.2-mm dust clumps throughout the Galactic plane. It became apparent upon inspection of the data presented in Table~\ref{tab:dust_sources} and Fig.~\ref{fig:long_lat_dust} that the water maser detection rate seemed to be higher for sources located at Galactic longitudes less than $\sim$G\,333. 

%\begin{figure}
%\begin{center}
%\vspace{-1cm}
%	\psfig{figure=lat_long_dust_1.eps,width=4.5cm,angle=270}\vspace{-1cm}
%	\psfig{figure=lat_long_dust_2.eps,width=4.5cm,angle=270}\vspace{-1cm}
%	\psfig{figure=lat_long_dust_3.eps,width=4.5cm,angle=270}
%	 \caption[Latitude versus longitude diagram of the targeted 1.2-mm dust clumps, showing the locations of the detected water masers.]{Latitude versus longitude diagram of the targeted 1.2-mm dust clumps. The 1.2-mm dust clumps with associated water masers are shown by crosses and those with no detectable water maser counterpart are represented by circles. The three 1.2-mm dust clumps which lie beyond $\pm$2 degrees of the Galactic plane have been excluded from these plots.}
%	 \label{fig:long_lat_dust}
%	 \end{center}
%\end{figure}

\subsection{Characteristics of the detected water masers}

%An extensive investigation of the characteristics of a large sample of water masers is carried out in \citet{Breen10b} and here we present a much less detailed investigation with our primary goal being to determine if the samples show any difference in their properties.

\subsubsection{Velocity ranges and flux densities of the complete sample}

The average velocity range of the 165 water masers is 26 \kms and they have a median velocity range of 13 \kmsns. In comparison, the average velocity range of the 379 water masers detected by \citet{Breen10b} is 29 \kms and they have a median velocity range of 15 \kmsns. While these numbers are quite similar, it is intriguing that velocity ranges of the sources that are potentially biased towards more evolved sources (\citet{Breen10b} chiefly targeted their observations towards OH masers) are, in general, larger than those detected towards the sample of dust clumps. \citet{Breen12stats} presents evidence that suggests that both the 6.7- and 12.2-GHz methanol maser sources show an increase in velocity range as the sources evolve. Although complicated by the presence of high-velocity features, we have carried out a similar analysis on the water masers in Section~\ref{sect:water_dust_ass}.

Further evidence that these water masers are less biased towards more evolved sources lies in the average and median flux density of the detected sources. \citet{Breen10b} present evidence that the water masers increase in flux density as the sources evolve. The average flux density of the water masers detected towards the dust clump sample is 39~Jy and they have a median of 4~Jy which are both lower than the average (57~Jy) and median (5~Jy) flux density of the sources presented in \citet{Breen10b}. This suggests that in the dust clump sample, we have detected water masers over a broader range of evolutionary stages.

\subsubsection{Association with other maser species and radio continuum}\label{sect:water_dust_ass}

\citet{Hill05} determined associations between their 1.2-mm dust clump sources and the 6.7-GHz methanol masers and \UCHII regions (detected at 8~GHz) towards which their observations were targeted. We have made an investigation of the number of dust clumps, with and without associated water masers, that \citet{Hill05} have designated as being: `mm-only' cores (i.e. sources without associated radio continuum emission or 6.7-GHz methanol masers); associated with 6.7-GHz methanol masers; associated with both 6.7-GHz methanol masers and radio continuum; and finally, those associated with only radio continuum. In all of our statistical analysis, 293 of the 294 targets for water masers have been included (omitting a single source which is missing values for some dust clump properties).

\begin{table}
\begin{center}
	\caption[Summary of the categories of dust clumps where we find water masers.]{Water maser detection rates in each dust clump category (from \citet{Hill05}): `mm-only' (those sources showing no associated methanol maser or radio continuum emission); associated methanol masers (meth); associated with both methanol masers and radio continuum associations (meth+cont); and associated continuum (cont). These numbers have been presented in two categories; those dust clumps with no detectable water maser emission, and those where we have detected water masers. The fourth column shows the water maser detection rate (per cent) in each of the categories.}
%	   \vspace{0.5cm}
\begin{tabular}{cccccc}\hline
 %& {\bf `mm-only'} & {\bf meth} & {\bf meth+cont} & {\bf cont} & {\bf total}\\ \hline \hline
 & {\bf No water} & {\bf Water} & {\bf Detection rate}\\
  &  &  & {\bf (\%)}\\ \hline \hline
 {\bf `mm-only'}		&	137	&	41	&	23\\
{\bf Meth}			&	13	&	48	&	79\\
{\bf Meth+cont}		&	5	&	25	&	83	\\
{\bf Cont}			&	10	&	14	&	58	\\ \hline
{\bf Total}			&	165	&	128	&	44	\\ \hline
% {\bf No water}	&	137	&	13	&	5	&	10	&	165\\
% {\bf water}	&	41	&	48	&	25	&	14	&	128\\ \hline 
         \label{tab:dust_water_ass}
	\end{tabular}
\end{center}
\end{table}

Table~\ref{tab:dust_water_ass} shows a summary of the categories of dust clumps where we detect the water masers, as well as the water maser detection rate in each category. We find an overall water maser detection rate of 44 per cent towards the targeted dust clumps, and this number falls to 23 per cent for the `mm-only' dust clump sources. The highest water maser detection rates are towards those dust clump sources that exhibit methanol maser emission, closely followed by those dust clumps showing only radio continuum emission. It can be seen that there are many more dust clump sources showing only water masers (41), than those showing only methanol masers (13). While some fraction of these 1.2-mm sources that show only methanol masers are likely to have associated water masers that were below our detection limits at the time of observation, it seems unlikely that this could be the case for all of them (by considering the statistics of detectability for the two epochs of water maser observations presented in \citet{Breen10b}). However, these sources will provide interesting targets for further water maser observations. \citet{Hill05} suggests that their `mm-only' cores may represent an earlier evolutionary phase than those sources showing methanol maser emission. If this is the case, these numbers certainly imply that water masers can be present even earlier than methanol masers. A supporting argument for water masers having a longer lifetime than that of methanol masers is that water molecules are both more easily created and more robust than methanol, which is consistent with our interpretation. 

An alternative explanation for the higher number of dust sources showing only water masers (than only methanol masers) is that a number of these additional water masers are associated with lower mass objects. Comparing the dust clump properties in each of the association categories offers no evidence for there being any systematic differences between the groups which may result if a large number of these sources were lower-mass objects.

We have investigated the water maser characteristics in each of the dust clump association categories. Table~\ref{tab:dust_water_flux_vel} presents the average and median values of both the water maser peak flux density and the velocity range. Like the methanol masers presented by \citet{Breen12stats}, we find that there is a general gradient which trends from lower values of both flux density and velocity range for the (probably) younger sources to higher values for the more evolved sources. This trend progresses through the categories as follows: those associated with `mm-only' sources, to those with methanol masers, and then to those with methanol masers and radio continuum \citep{Breen10a,Hill05}. The exception to this trend is the last stage, where only radio continuum is seen.

\begin{table*}
\begin{center}
	\caption[The average and median flux densities, and velocity ranges of all of the detected water masers, broken into the different association categories]{The average and median flux densities, and velocity ranges of all of the detected water masers. Water maser sources have been broken up into several categories, and are as for Table~\ref{tab:dust_water_ass}, with the addition of `all sources' which shows the values for the entire sample. }
%	   \vspace{0.5cm}
\begin{tabular}{cccccc}\hline
 {\bf Water }& {\bf Average } & {\bf Median } & {\bf Average} & {\bf Median } \\
 {\bf classification}& {\bf flux density} & {\bf flux density} & {\bf  vel range} & {\bf vel range} \\
&				{\bf (Jy)}	&	{\bf (Jy)}	&	{\bf (\kmsns)}	&   {\bf (\kmsns)}\\ \hline \hline
{\bf all sources}		&	39 	&	4.4	&	26	&	13	\\
{\bf `mm-only'}		&	2	&	1.5	&	19	&	8\\
{\bf Meth}			&	44	&	9	&	28	&	19\\
{\bf Meth+cont}		&	107	&	18.5	&	42	&	23	\\
{\bf Cont}			&	9	&	1	&	26	&	17	\\ \hline
         \label{tab:dust_water_flux_vel}
	\end{tabular}
\end{center}
\end{table*}

%## cut this figure 25 Jan
%\begin{figure}
%\begin{center}
%	\psfig{figure=lum_vrange.eps,width=6cm,angle=270}
%	 \caption[]{Log of the integrated water maser luminosity versus the water maser velocity range. Shown on the plot are water sources that are associated with `mm-only' sources (purple dots); water masers associated with methanol masers (black triangles); water masers associated with both methanol masers and 8~GHz radio continuum (red crosses); and water masers associated with only 8~GHz radio continuum (green stars). Overlaid is the line of best fit that corresponds to all of the water masers presented on the plot (linear regression gives; slope=12.11[2.37], {\em p}-value= $<$ 1.64e-06, and a correlation coefficient of 0.45).}
%	 \label{fig:water_dust_vrange_lum}
%	 \end{center}
%\end{figure}
%

%Call:
%lm(formula = vrange ~ lum, data = a)

%Residuals:
%    Min      1Q  Median      3Q     Max 
%-35.458 -13.872  -7.680   6.743 122.207 

%Coefficients:
%            Estimate Std. Error t value Pr(>|t|)    
%(Intercept)   -4.034      6.195  -0.651    0.516    
%lum           12.111      2.373   5.103 1.64e-06 ***
%---
%Signif. codes:  0 Ô***Õ 0.001 Ô**Õ 0.01 Ô*Õ 0.05 Ô.Õ 0.1 Ô Õ 1 

%Residual standard error: 25.57 on 98 degrees of freedom
%Multiple R-Squared:  0.21,	Adjusted R-squared: 0.2019 
%F-statistic: 26.04 on 1 and 98 DF,  p-value: 1.637e-06 

If the numbers presented in Table~\ref{tab:dust_water_flux_vel} are representative of the larger population of water masers, then they imply that, while the water masers increase in flux density and velocity range as the sources evolve, towards the end of their lifetime this trend `turns over' and they steadily decrease in both flux density and velocity range, presumably decreasing until the maser emission ceases. This is contrary to that seen in the case of methanol masers \citep{Breen12stats}, which seem not to experience any downwards trend, implying that they switch off much more abruptly.

Comparing the luminosities of the water masers (calculated using the near distances presented in \citet{Hill05}) with the velocity ranges of the water maser sources, similarly shows an increase in both luminosity and velocity range through the different association categories from least to most evolved. However, there is significant overlap between the values within each association category. A high level of scatter is expected when considering water masers, since a number of sources show high velocity features and it is not clear that these are restricted to a certain evolutionary stage. Also, the sources that are associated with only 8~GHz radio continuum are scattered throughout the range of values that are associated with the sources in other categories, but none have large values of luminosity, and very few have large velocity ranges. These findings are consistent with those conclusions drawn from the average values that are presented in Table~\ref{tab:dust_water_flux_vel} where peak flux densities are considered.

%## cut this para 25 Jan
%In Fig.~\ref{fig:water_dust_vrange_lum}, the luminosities and velocity ranges of the water masers are presented, broken into the same categories of associations as in Table~\ref{tab:dust_water_ass} and Table~\ref{tab:dust_water_flux_vel}. It can be seen that there is a steady increase in both water maser luminosity and velocity range as you move through the association categories; `mm-only'; to those with associated methanol masers; to those with both methanol masers and radio continuum. However, it is evident that there is significant overlap between the values within each association category. A high level of scatter is expected when considering water masers, since a number of sources show high velocity features and it is not clear that these are restricted to a certain evolutionary stage. Also shown on the plot are the sources that are associated with only 8~GHz radio continuum and these sources are scattered throughout the plot but none have large values of luminosity, and very few have large velocity ranges. These findings are consistent with those conclusions drawn from the average values that are presented in Table~\ref{tab:dust_water_flux_vel}. Linear regression analysis of the full set of water masers (i.e. all of the association categories combined) shows that there is a statistically significant positive slope in the line of best fit and that there is a moderate correlation between data points.

%22-GHz sources younger? that is the reason for the difference between this and chapter whatever?

\subsubsection{Comparison between water maser luminosity and 1.2-mm dust clump H$_2$ number density}
\begin{table}
\begin{center}
	\caption[Relationship between 6.7 and 12.2-GHz methanol and water maser luminosities and associated H$_2$ number densities of the 1.2-mm dust clumps]{Slopes and intercepts of the line of best fit of log integrated maser luminosity versus log of the H$_2$ number densities of the 1.2-mm dust clumps. Values for 6.7- and 12.2-GHz methanol masers are listed for comparison with the water maser values for comparison. Errors are presented in square brackets and follow the estimates of each value.}
%	   \vspace{0.5cm}
\begin{tabular}{cccccc}\hline
 {\bf Maser}& {\bf Slope} & {\bf Intercept } & {\bf Reference} \\
{\bf species}&				 \\ \hline \hline
{\bf 6.7-GHz}		&	--0.72[0.25] 	&	5.53[1.17]		& \citet{Breen12stats}\\
{\bf 12.2-GHz}		&	--0.56[0.28]	&	4.16[1.23]		& \citet{Breen10a}\\
{\bf water}			&	--0.43[0.07]	&	3.86[0.33]		& current \\\hline
         \label{tab:density_lum_masers}
	\end{tabular}
\end{center}
\end{table}

%log(12.2 GHz lum) = ?0.56[0.28](log(H2 density)) + 4.16[1.23], - integrated
%log(6.7GHzlum)= ?0.85[0.16](log(H2density))+6.25[0.75], - peak
%log(6.7 GHz lum) = ?0.72[0.25](log(H2 density)) + 5.52[1.17], -integrated

%## Cut this figure 25th Jan
%\begin{figure}
%\begin{center}
%	\psfig{figure=lum_density.eps,width=9cm,angle=270}
%	 \caption[]{Log water maser peak luminosity versus log of the H$_2$ number density of the associated 1.2-mm dust clump. The dashed line shows the line of best fit (which has a slope of --0.13[0.05] and an intercept of 4.79[0.11] and a correlation coefficient of 0.21).}
%	 \label{fig:lum_density_water_dust}
%	 \end{center}
%\end{figure}

%Call:
%lm(formula = density ~ lum, data = s)

%Residuals:
%     Min       1Q   Median       3Q      Max 
%-1.21691 -0.39991 -0.04078  0.29369  2.15533 

%Coefficients:
%            Estimate Std. Error t value Pr(>|t|)    
%(Intercept)  4.78817    0.11127  43.032  < 2e-16 ***
%lum         -0.13208    0.04967  -2.659  0.00884 ** 
%---
%Signif. codes:  0 Ô***Õ 0.001 Ô**Õ 0.01 Ô*Õ 0.05 Ô.Õ 0.1 Ô Õ 1 

%Residual standard error: 0.5618 on 126 degrees of freedom
%Multiple R-Squared: 0.05315,	Adjusted R-squared: 0.04563 
%F-statistic: 7.072 on 1 and 126 DF,  p-value: 0.008845 

\begin{figure}
\begin{center}
	\psfig{figure=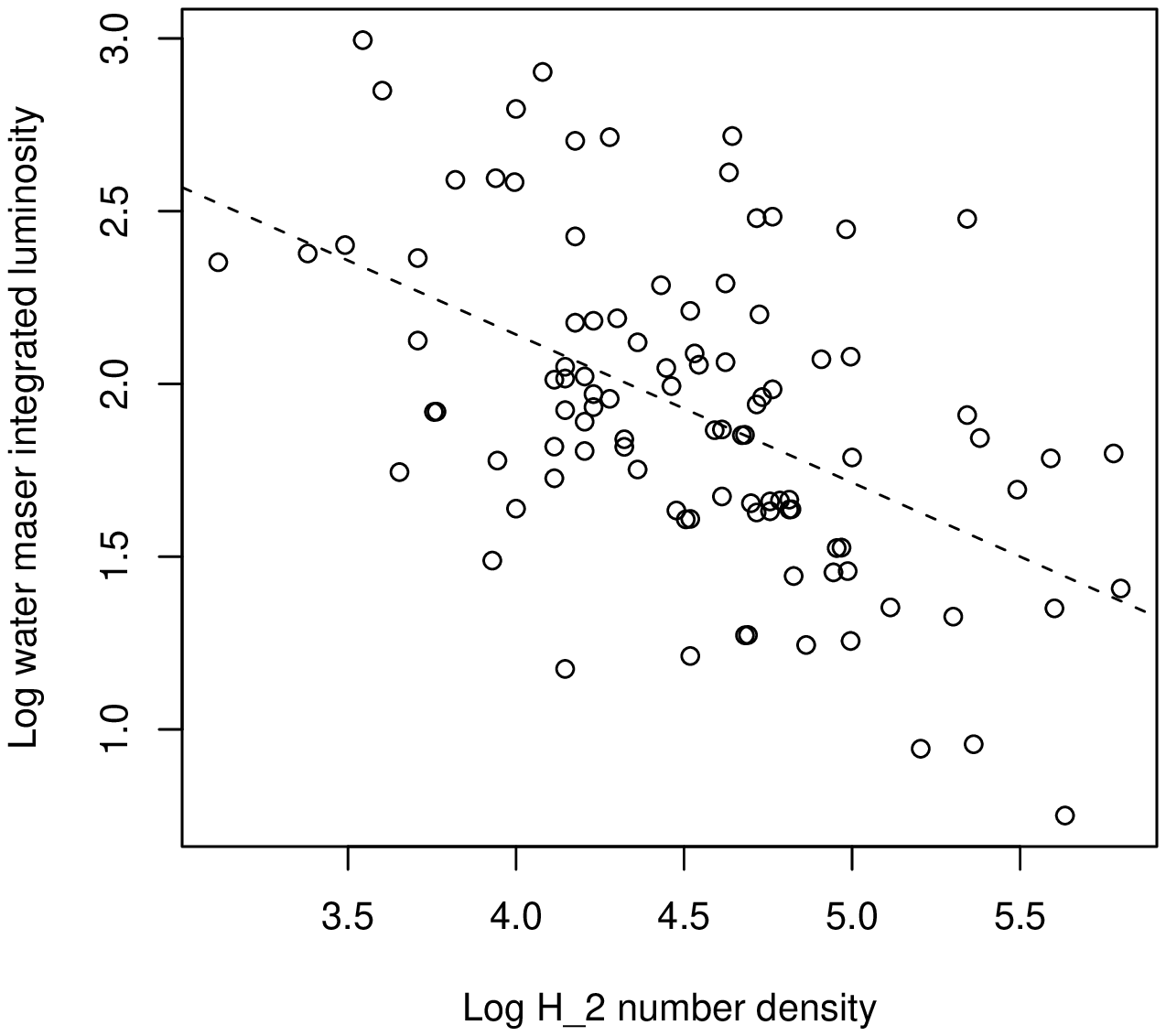,width=9cm,angle=270}
	 \caption[]{Log water maser integrated luminosity (Jy \kms kpc$^2$) versus log of the H$_2$ number density (cm$^{-3}$) of the associated 1.2-mm dust clump. The dashed line shows the line of best fit (which has a slope of --0.43[0.07] and an intercept of 3.86[0.33] and a correlation coefficient of 0.50).}
	 \label{fig:integ_lum_density_water_dust}
	 \end{center}
\end{figure}

%
%Call:
%lm(formula = lum ~ density, data = s)

%Residuals:
%     Min       1Q   Median       3Q      Max 
%-0.90522 -0.25201 -0.04934  0.29091  0.91027 

%Coefficients:
%            Estimate Std. Error t value Pr(>|t|)    
%(Intercept)  3.85756    0.33111  11.650  < 2e-16 ***
%density     -0.42864    0.07268  -5.898 5.26e-08 ***
%---
%Signif. codes:  0 Ô***Õ 0.001 Ô**Õ 0.01 Ô*Õ 0.05 Ô.Õ 0.1 Ô Õ 1 

%Residual standard error: 0.3922 on 98 degrees of freedom
%Multiple R-Squared: 0.2619,	Adjusted R-squared: 0.2544 
%F-statistic: 34.78 on 1 and 98 DF,  p-value: 5.259e-08 

\citet{Breen10a} shows that the H$_2$ number density of the 1.2-mm dust clumps decreases as the sources evolve. Although, as noted by \citet{Breen10a}, the apparent change in density might be a consequence of the constant temperature assumption applied by \citet{Hill05} in their calculations and consequently the difference in densities might instead represent an increase in temperature. In either case, the changes to the physical properties of the source are consistent with evolution.

 In Fig.~\ref{fig:integ_lum_density_water_dust} we present the log of the integrated water maser luminosity versus the log H$_2$ number densities of the associated 1.2-mm dust clumps. This plot show a slope that is consistent with the findings of \citet{Breen10a} and those previously presented in this section - that the more luminous water masers are more evolved and that these are associated with the less dense 1.2-mm dust clumps and vice versa. Table~\ref{tab:density_lum_masers} compares the computed slopes and intercepts of the line of best fit of the log luminosity 6.7- and 12.2-GHz methanol masers, along with the water masers versus the H$_2$ number densities of the associated clumps.

Inspection of the values listed in this table show firstly (as previously stated) that the 6.7-GHz methanol masers increase in flux density more rapidly as they evolve than the associated 12.2-GHz sources, and, as shown by the larger intercept, are generally stronger than the 12.2-GHz sources. Interestingly, the values associated with the water maser sources imply that the water masers also increase in flux density less rapidly than the 6.7-GHz methanol masers as they evolve and are also weaker in general at a given H$_2$ number density (than the sample of 6.7-GHz methanol masers). However, it is difficult to determine which of the maser species are present at the earliest evolutionary stage of high-mass star formation and therefore which species shows stronger emission at a given evolutionary stage. %While the water masers seem to be weaker than the 6.7-GHz methanol masers at a particular evolutionary stage, it is possible that the water masers spend a significant fraction of their lifetime as relatively weak sources. 

\subsubsection{Water maser variability}

Water masers are known to exhibit extreme variability over relatively short timescales \citep[e.g.][]{Felli07}. The water maser observations presented in \citet{Breen10b} were completed, in many cases, twice with the observation epochs separated by $\sim$10 months. Additionally, \citet{CB10} carried out a double epoch search for water maser emission of two small regions of the Galactic plane. Since the observations presented here were completed only once, we cannot directly determine the level which maser variability has affected our results. Instead, we derive predictions about the completeness of these observations by comparing this sample with the findings of \citet{Breen10b} and \citet{CB10}.

\citet{Breen10b} finds that 17 per cent of the 253 water maser sources that were observed at two epochs were only detectable at one of these epochs. \citet{CB10} found similarly that 16 per cent of the 32 sources that they detected in their double epoch complete searches were detectable only once. If we simply apply the larger percentage (i.e. 17 per cent) to our current sample, it might be expected that $\sim$34 additional water maser sources would be detected if a second epoch of observations were conducted towards the target 1.2-mm dust clumps. Since we detect 165 water masers towards 128 1.2-mm dust clumps, we would perhaps expect that these additional sources would be detected towards a further 26 1.2-mm dust clumps. 

However, the situation is a little more complex than this. It is not clear if the role of variability would result in many more dust clumps being recognised as being associated with water masers, or alternatively that different water maser sources would be detected at different epochs towards the same dust clump. It is likely that both would be true, that is, that some number of water masers would be detected towards additional 1.2-mm dust clumps and that some additional/different water masers would be detected towards 1.2-mm dust clumps where we have detected water masers.

\subsection{Location of the water masers in the 1.2-mm dust clumps}\label{sect:water_dust}

Analysis of the locations of the water masers with respect to the 1.2-mm dust clump peak, as reported by \citet{Hill05}, shows that there is excellent correspondence between their locations. The 1.2-mm dust continuum maps of \citet{Hill05} have 8 arcsec pixels, and the listed peak positions are just the central position of the peak pixel. On average, we find that the separation between the water maser positions and the associated dust clump peak is 13 arcsec and that the median separation is 8 arcsec. It is therefore likely that, in the majority of cases, the water masers are very closely associated with the peak 1.2-mm dust clump emission.

\begin{figure}
\begin{center}
\vspace{-1cm}
\psfig{figure=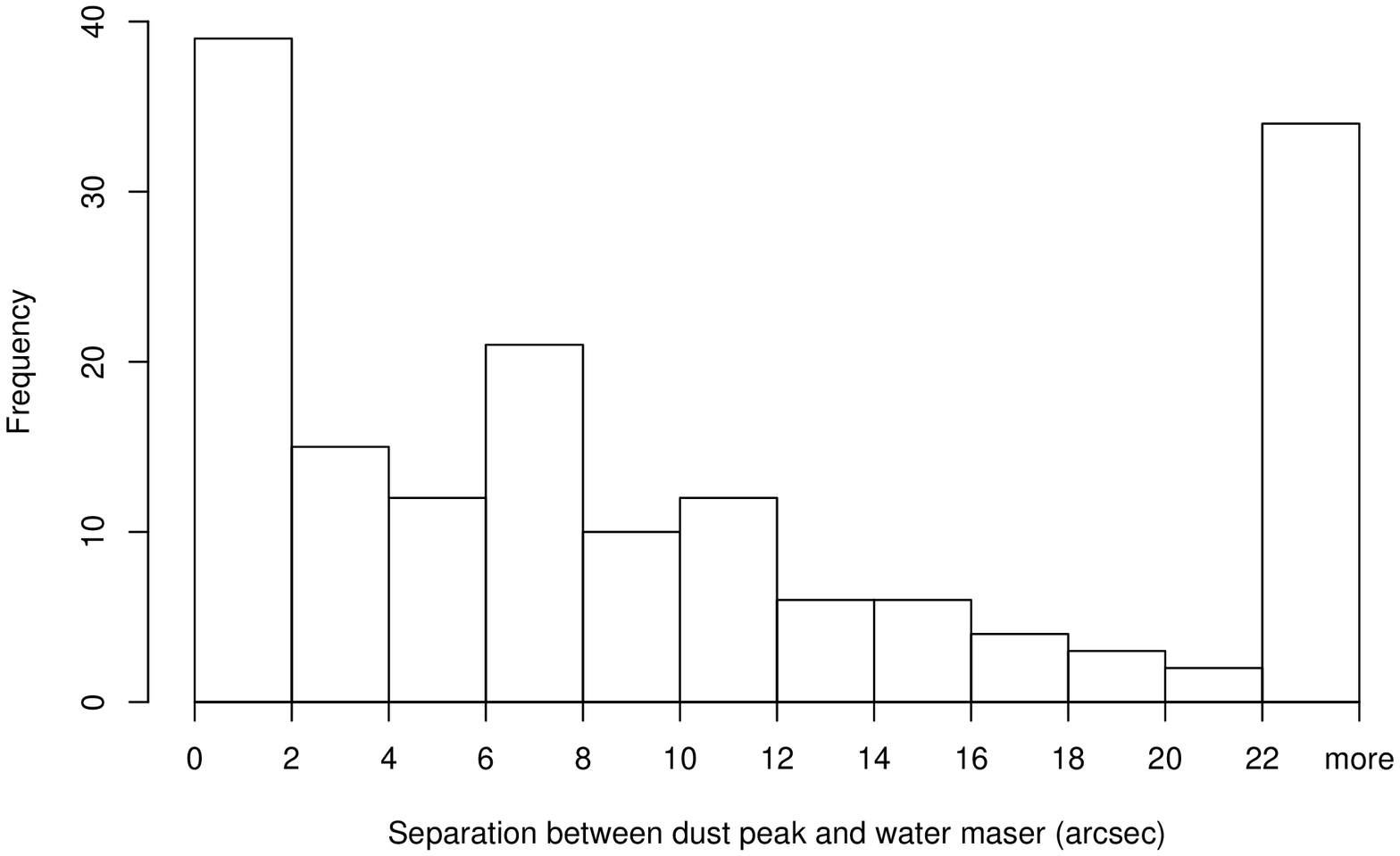,width=6cm,angle=270}
	 \caption{Histogram of the angular separation between the reported 1.2-mm dust clump peak and the position of the water maser sources. }
	 \label{fig:sep_hist}
	 \end{center}
\end{figure}

Fig.~\ref{fig:sep_hist} shows a histogram of the separations between the dust clump peak and the detected water masers. All water masers that were separated by more than $\sim$30 arcsec from the dust clump peak were further investigated by comparing the reported dust clump peak position with the dust maps presented in figure A1 of \citet{Hill05} and the water maser positions. We find in just over half of these cases (11 of 19) there is an obvious error in the reported dust clump position which has lead to an apparently large separation from the water maser position. In the worst examples, the reported dust clump peak is located beyond the boundary of the respective clump. Even though \citet{Hill05} publish images of their 1.2-mm dust clumps, the resolution (and contrast) is insufficient to enable us to determine better positions independently. Given the obvious problems with some number of the dust clump peak positions which erroneously result in some of the most extreme separations between the peak of the clump and the associated water maser, it is likely that the actual average and median separations between the peak of the dust clumps and the associated water masers would be even smaller, perhaps comparable to the pixel size. 

%For example, dust clump G\,330.95--0.18 has a reported position that is 73 arcsec from the location of the water maser, but inspection of the 1.2-mm dust clump map shows that the maser is actually very close to the peak. In this case, the reported dust clump peak is outside the visible portion of the clump. There are numerous examples where a similar thing is seen (e.g. dust clumps G\,294.97--1.7, G\,299.02+0.1 and G\,301.14--0.2), but a more thorough investigation is needed to determine exactly how many sources are affected. 

We have compared the locations of the water masers with respect to the location of their associated dust clump peak and the peak flux density of the maser sources. It is evident that very few of the water masers that are truly significantly offset from the peak of the dust clumps have flux densities that are greater than a few~Jy. The possible corollary, that most of the water masers that are located far from the dust clump peak are relatively weak, needs further investigation.

%## Cut 25 Jan
%Fig.~\ref{fig:location_flux} shows a plot of the separation from the dust clump peak versus the peak flux density of the associated water masers. This figure shows, similarly to Fig.~\ref{fig:sep_hist}, that the bulk of the water maser sources are located  near the peak of the associated 1.2-mm dust clumps. No strong trend is seen between the peak flux density and the separation between the dust clump peak position and the peak flux density of the associated water maser, but investigation of the 10 strong water masers that are located $>$16 arcsec from the dust clump peak shows that many of these have erroneously reported dust clump peak positions. Further investigation is therefore required, however, it is possible that most of the water masers that are located far from the dust clump peak are relatively weak.

%### removed 25 Jan	
%\begin{figure}
%\begin{center}
%\vspace{-1cm}
%	\psfig{figure=location_flux.eps,width=6cm,angle=270}
%	 \caption[Separation from dust clump peak versus water maser flux density plot.]{Separation between the dust clump peak and the water masers associated with each of the dust clumps versus the water maser peak flux density. The vertical dashed line is marked at 16 arcsec. }
%	 \label{fig:location_flux}
%	 \end{center}
%\end{figure}
%	

\subsection{Assessing the current model of water maser presence in 1.2-mm dust clumps}

\citet{Breen} created a model for water maser presence towards 1.2-mm dust clumps which was derived from observations of the 1.2-mm dust continuum emission \citep{Mook04} and water maser observations that they completed within the G333 giant molecular cloud. By determining which 1.2-mm dust clumps were associated with water masers as well as those that were devoid of water maser emission, \citet{Breen} fitted a Binomial Generalized linear model to the maser presence/absence data, using the 1.2-mm dust clump properties as predictors. They found that the simplest model with the greatest predictive properties was based only on dust clump radius.

We have used the model derived by \citet{Breen} to calculate the probability of water maser detection for each of the dust clumps presented in \citet{Hill05} that have dust clump radius measurements. For a number of dust sources presented in \citet{Hill05} there is a near-far distance ambiguity. In these cases, we have assumed the near distance for consistency with analysis completed by \citet{Breen10a}. However, we note that in this instance this assumption might be particularly hazardous, since the predictive model relates only to 1.2-mm dust clump radius which is highly influenced by the adopted distance measurement. 

Using the model for water maser presence associated with 1.2-mm dust clumps, it was determined that only 58 of the 294 dust clumps considered in this analysis, had a probability of 0.01 or greater of having an associated water maser. Probabilities for individual dust clumps are presented in Table~\ref{tab:dust_sources}. Comparison between the calculated probability of water maser and the actual detections shows that the model described in \citet{Breen} is promising. The average probability of water maser presence in the 128 1.2-mm dust clumps where water masers are detected is 0.10, and has a median value of 0.00081. In comparison, for the dust sources where we detect no water maser emission, the average probability of water maser presence is 0.027 and has a median value of 0.00026. An alternative description of these statistics is that, of the 27 1.2-mm dust clumps which had a calculated probability of $\geq$0.1, 20 yielded water maser detections. If the considered probability is lowered to be $\geq$0.01 then the number of predicted clumps is increased to 58 and the number of clumps with detections is 39. It is therefore clear that the model is picking up a high detection rate for the sources with high calculated probabilities, however, given that we detect water masers towards 128 dust clumps, it is also clear that the model is failing to return high probabilities of water maser presence in around two thirds of the clumps with detections.

Given that the original model was derived from a sample of water masers that were detected in a complete search with a relatively poor sensitivity limit ($\geq$5~Jy), the obvious property to investigate is the relative flux density of the detected water masers. It might be expected that there would be a trend whereby the water maser detections towards the 1.2-mm dust clumps with high association probabilities have higher flux densities, given the nature of the data on which the model was derived. The average flux density of the water maser sources detected towards dust clumps with a probability of 0.01 or higher (58 water maser sources) is 48~Jy and these sources have a median flux density of 6~Jy. In comparison, the average flux densities of the water masers detected towards clumps with a lower calculated probability is 33~Jy, with a median flux density of 3.5~Jy. It is possible, therefore, that the current model is biased towards predicting correctly the presence of the stronger sources. However, more likely, is that the sources with very low probabilities are dominated by far distance sources which accounts for the lower average flux density (which would also mean that their probabilities have been under estimated).

Another obvious factor to consider is the high degree of dependence on distance measurements. Since the original model was derived from sources located within the one GMC with a well constrained distance measurement, the poorly determined distances towards most star formation regions did not affect the generation of this model. The model for water maser presence associated with dust clumps uses only the radius of the dust clump in order to calculate the probability. As previously discussed, we have assumed the near distance reported by \citet{Hill05} for sources with distance ambiguities. It is clear that this is an unreliable assumption, especially since our search is of high sensitivity which reduces the likelihood of any significant detection dependence on distance. It would be expected that the split of near and far distance allocations should be approximately even (or maybe even favour the far distances due to the large volume). Inspection of the characteristics of some of the dust clump sources that have low calculated probabilities, yet have associated water masers, shows that if the far distance was used the calculated probability would be high (in many cases $>$0.3). The increase in radius that results in a change from near to far distance does not result in a high probability in water maser presence in all of the dust clump sample, especially those with no detected associated water masers. %higher detection rate for nearby sources? Should look at the distances... is there a sensitivity limitation?

If our assertion, that the poorly constrained distance measurements are adversely affecting the predictive capabilities of the model, is correct, then the distribution of the clump properties should support this. If, for example, we inspect the peak flux density of the 1.2-mm dust clumps with associated water masers and calculated probabilities of $>$ 0.01, they should not be any different from the sample of 1.2-mm dust clumps with associated water masers and calculated probabilities of $<$0.01. But, they are in fact different. The average peak flux density of the clumps with calculated probabilities of $>$0.01 (and have associated water masers) is 2.8 Jy beam$^{-1}$, whereas the average of those clumps with lower than expected calculated probabilities ($<$0.01) is 1.4 Jy beam$^{-1}$. Again, the most likely explanation for this is that a number of the 1.2-mm dust clumps which have low calculated probabilities, but yet have an associated water maser, are located at the far distance. This means that the seeming downfall of the model to allocate high probabilities to a number of the 1.2-mm dust clumps where we detect water masers may be eradicated if accurate distance measurements were obtained.

%Even though there is obvious scope to improve the model of \citet{Breen}, it is showing greater potential than other models for water maser presence. 

%clustering of high prob non detections - distance shit or maybe something else going on??

%what is the integrated and peak fluxes of the sources with a high prob that have and don't have actual water masers????

%is there are trend between water maser luminosity and clump anything?

\subsection{Fitting Binomial general linear models to the new water maser and 1.2-mm dust clump data}		   

While the current model for water maser presence in 1.2-mm dust clumps is promising, we have repeated the analysis carried out by \citet{Breen} with the much larger sample presented here. However, two important considerations are: firstly, the adoption of the near distance measurements for the sources with distance ambiguities; and secondly, that the 1.2-mm dust clumps in this sample were not detected in a systematic manner (unlike the observations of \citet{Mook04} used in \citet{Breen}), although given the high number of additional 1.2-mm dust sources found within the fields of the target sources this is probably not detrimental to the analysis. 

We have fitted a Binomial generalized linear model to the water maser presence/absence data, using the 1.2-mm dust clump properties as predictors. For a more detailed description of this analysis method, see \citet{Breen}. All but one of the targeted 1.2-mm dust clumps were included in this analysis. The source that has been excluded is G\,6.60--0.08 as  \citet{Hill05} suggest that the derived mass of this source is uncharacteristic of high-mass star formation regions. For each of the 1.2-mm dust clumps, all of the derived clump properties were tested: the integrated flux density (Jy), peak flux density (Jy beam$^{-1}$), source full width at half maximum (FWHM) (arcsec), distance (kpc), mass (M$_{\odot}$), radius (pc) and H$_{2}$ number density (cm$^{-3}$). In the following sections we consider {\em p}-values of less than 0.05 to be statistically significant (i.e. the hypothesis that the single term model provides no better fit than the null model consisting only of an intercept is rejected when the {\em p}-value is less than 0.05).

\begin{figure}
\begin{center}
	\psfig{figure=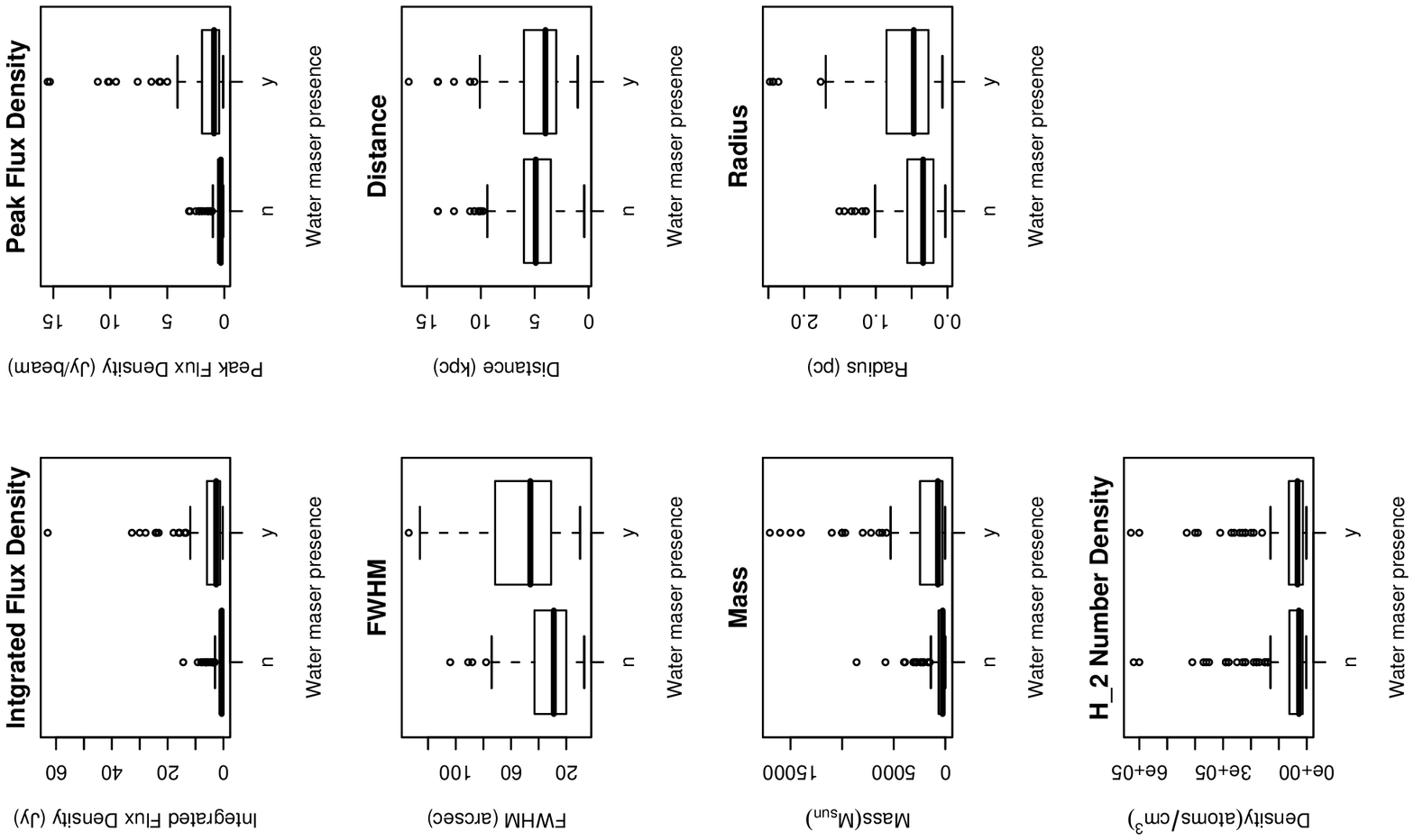,width=15cm,angle=270}
	 \caption[Box plots of the 1.2-mm dust clump properties split in the categories of y and , referring to the presence of water maser emission. ]{Box plots of the 1.2-mm dust clump properties split in the categories of yes and no, referring to the presence of water maser emission. There is a statistically significant difference between the two categories in the 1.2-mm dust clump integrated and peak flux density, mass, FWHM and radius. This information is the graphical display relating to the information in Table~\ref{tab:dust_single}. Note that source G\,23.960+0.137 has been removed from these plots, since is is more than twice as dense as all of the other sources shown on the plot and therefore is an extreme outlier.}
	 \label{fig:water_dust_box}
	 \end{center}
\end{figure}

Firstly, all dust clump properties were tested to see if individually they could give an indication of the likelihood of associated water maser presence. This was done by fitting a single term addition Binomial model to each property (or 'predictor') and showed an increasing probability of the presence of  water masers associated with increasing values of dust clump integrated flux density, peak flux density, FWHM, mass and radius (shown in Table~\ref{tab:dust_single}). This result is similar to that found when investigating methanol masers \citep{Breen10a}, and means that any one of these dust clump properties can give an indication of the likelihood of finding an associated water maser. Box plots of each of the dust clump properties broken up into the categories of `n' and `y', referring to those clumps with no associated water maser, and those with an associated water maser, respectively, are presented in Fig.~\ref{fig:water_dust_box}. These box plots show graphically the same information as the results of the single term additions; that lower values of the dust clump properties; integrated and peak flux density, FWHM, mass and radius, are associated with the 1.2-mm dust clump sources with no associated water maser emission, compared to those associated with water maser emission.

\begin{table}
\begin{center}
  \caption[Analysis of deviance table for the single term models (using the 1.2-mm dust clump properties from \citet{Hill05}).]{Analysis of deviance table for the single term models (using the 1.2-mm dust clump properties from \citet{Hill05}), showing the deviance and the AIC together with the associated likelihood ratio statistic and {\em p}-value for the test of the hypothesis that the stated single model provides no better fit than the null model consisting only of an intercept.}
  %   \vspace{0.5cm}
\begin{tabular}{ccccc}\hline
	{\bf Predictor} & {\bf Deviance} & {\bf AIC} &  {\bf LRT} & {\bf {\em p}-value}\\ \hline  \hline
{\bf none} &    399.84 & 401.84    \\                   
{\bf integ} &      341.81 &345.81 & 58.02 &2.590e-14\\ 
{\bf peak}   &    332.98 &336.98&  66.85& 2.923e-16 \\
{\bf fwhm}    &   354.23 & 358.23 & 45.61 & 1.445e-11\\ 
{\bf dist}       & 398.65 &402.65 &   1.19&    0.2761    \\
{\bf mass}     &   366.41 &370.41  &33.43 &7.391e-09\\
{\bf radius}     & 379.44 &383.44 & 20.40& 6.299e-06\\ 
{\bf density}   & 399.66 &403.66&   0.18 &   0.6697\\     \hline      
         \label{tab:dust_single}
	\end{tabular}
\end{center}
\end{table}

After determining the predictive capabilities of individual dust clump properties, stepwise model selection was used to create the simplest model with the greatest predictive power. This method considers all of the dust clump properties and tries to maximise the predictive properties while trying to minimise the number of terms (and therefore dust clump properties). The resultant model contains only two dust clump properties; peak flux density and FWHM. The estimated regression relation is

\begin{eqnarray*}
  \mbox{log}{\frac{p_{i}}{1-p_{i}}}=-1.564+0.995x_{\rm Peak}+0.012x_{\rm FWHM},
\end{eqnarray*}
where {\em x}$_{\rm Peak}$ is the peak flux density, {\em x}$_{\rm FWHM}$ is the source FWHM, and {\em p$_{i}$} is the probability of finding a water maser towards the {\em i$^{th}$} 1.2-mm dust clump. The regression summary of this model is shown in Table~\ref{tab:dust_reg}. In contrast to the previous dataset where a model based on dust clump radius (pc) was found to be best, the current dataset uses dust clump peak flux density and FWHM (arcsec). As can be seen in Table~\ref{tab:dust_reg}, the dust clump peak flux density is the most influential term, indicated by the large standard error relative to the coefficient of the dust clump FWHM together with the large {\em p}-value.

\begin{table}
\begin{center}
	\caption[Summary table for the Binomial regression model.]{Summary table for the Binomial regression model, showing for each predictor the estimated coefficient and the standardised z-value and {\em p}-value for the test of the hypothesis that $\beta_{i}$=0.}
%	   \vspace{0.5cm}
\begin{tabular}{ccccc}\hline
	{\bf Predictor}& {\bf Estimate} & {\bf Std. Error} & {\bf z value} & {\bf {\em p}-value}\\ \hline \hline
{\bf Intercept}& -1.563746   &0.271202 & -5.766 & 8.12e-09\\
{\bf peak}        & 0.994999 &  0.254413 &  3.911 & 9.19e-05 \\
{\bf fwhm}        & 0.011744 &  0.007242 &  1.622 &   0.105\\ \hline
         \label{tab:dust_reg}
	\end{tabular}
\end{center}
\end{table}

The misclassification rates for this new model are somewhat promising when the probability threshold is set to 0.5, it correctly predicts half (63 of 127) of the dust clump sources with associated water masers, and it correctly predicts 144 of the 166 % this previously said 165 which means that I was short 1 dust source
dust clumps with no associated water masers. It is easy to understand why such a model would fail on a high number of the dust clumps with associated water masers, since we know that these sources have higher peak flux densities on average, but as can be seen in Fig.~\ref{fig:water_dust_box}, there is a large overlap in the range of values of this property between those clumps with associated water masers and those without. This is consistent with a number of dust clumps with associated water masers having apparently low peak flux densities, since they are located at the far distance. Therefore a peak flux density measurement that is scaled with reliable distance measurements (a pseudo luminosity) may offer a promising model. Likewise, a reliable model may be derived from radius measurements that have correctly accounted for distances.

%luminosity would be better 
%    FALSE TRUE
%  n   144   21
%  y    64   63

\subsubsection{A new model based on 1.2-mm dust clump radius}

Since the model for water maser presence presented in \citet{Breen} uses only the dust clump radius to predict the probability of associated water maser presence, for comparison, a model using only dust clump radius has been produced from this new data. The resultant estimated regression relation is %This was achieved by restricting the dust clump properties used to fit the GLM to only the dust clump radius. 

\begin{eqnarray*}
  \mbox{log}{\frac{p_{i}}{1-p_{i}}}=-0.981+1.45x_{\rm radius},
\end{eqnarray*}
where {\em x}$_{\rm radius}$ is the radius of the 1.2-mm dust clump, and {\em p$_{i}$} is the probability of finding a water maser towards the {\em i$^{th}$} 1.2-mm dust clump. The regression summary of this model is shown in Table~\ref{tab:dust_reg_radius}.

\begin{table}
\begin{center}
	\caption[Summary table for the Binomial regression model using only dust clump radius]{Summary table for the Binomial regression model using only dust clump radius, showing the estimated coefficient and the standardised z-value and {\em p}-value for the test of the hypothesis that $\beta_{i}$=0.}
%	   \vspace{0.5cm}
\begin{tabular}{ccccc}\hline
	{\bf Predictor}& {\bf Estimate} & {\bf Std. Error} & {\bf z value} & {\bf {\em p}-value}\\ \hline \hline
{\bf Intercept}& -0.9814  &   0.2101 &  -4.670 & 3.01e-06 \\
{\bf radius}     & 1.4460  &   0.3544  & 4.081 &4.49e-05  \\ \hline
         \label{tab:dust_reg_radius}
	\end{tabular}
\end{center}
\end{table}

Since only one dust clump property is present in the regression relation, it is easy to determine the physical implications of the model. For example, if the probability of water maser presence is set to 0.5, the corresponding dust clump radius is 0.68 pc. This means that the model is saying that 1.2-mm dust clumps with a radius of 0.68 pc (or higher) have a probability of 0.5 (or higher) of having an associated water maser. In comparison, the model presented by \citet{Breen} predicts that 1.2-mm dust clumps with radii greater than 1.25 pc for the same probability threshold of 0.5 will have associated water masers. The model produced with the current data, predicts that water masers are associated with much smaller dust clumps (about half the size; which is perhaps not surprising given that these observations are of much higher sensitivity), albeit relatively poorly. The misclassification rates from this new model are poor, particularly in determining which clumps will have associated water masers (which it gets wrong two-thirds of the time).

The poorly constrained distance measurements are therefore having a large effect on our ability to model the water maser presence within the 1.2-mm dust clumps. However, the higher accuracy of the models in predicting the sources without associated water masers correctly, may indicate that there is a population of sources that are significantly far from the properties of the clumps that exhibit water masers that not even a change from near to far distance would boost them to a high probability of water maser presence. Therefore, from the current data we have been unable to derive an adequate model for water maser presence from dust clump radius.

%Both models are ok at predicting the sources that will not have masers... - need to think about that... probs just means that the ones that wont have even smaller values than the ones that are wrongly assigned to the near distance..

%
%    FALSE TRUE
%  n   137   28
%  y    84   43

\section{Conclusion}

%A water maser search towards 267 of the 1.2-mm dust clump catalogue presented by \citet{Hill05} has resulted in the detection of 132 water masers towards 101 dust clumps. 

We find water maser emission towards 128 of the 294 1.2-mm dust clumps searched. In total, we detect 165 distinct water maser sites and most are either new, or achieve accurate positions for the first time.

There is an excellent correspondence between the positions of the water masers and the peaks of the 1.2-mm dust clumps in the majority of sources. In addition to the water masers that we detect towards our target 1.2-mm dust clumps, we detect four sources towards regions apparently devoid of dust continuum emission (from comparison between the water maser locations and the 1.2-mm continuum maps of \citet{Hill05}). We suggest that two of these sources may be located at the far distance which has resulted in lower sensitivity 1.2-mm continuum observations towards these sources.

\citet{Hill05} allocated their 404 1.2-mm dust clumps in their catalogue the following categories: associated with a 6.7-GHz methanol maser; associated with both a methanol maser and 8-GHz radio continuum; associated with radio continuum only; and millimetre only, i.e. not associated with either methanol masers or radio continuum. We have assessed our detection rates of water masers towards the dust clumps in each of these categories. The highest water maser detection rates are towards dust clumps which are associated with methanol masers (both with and without radio continuum), and the lowest detection rate is towards the millimetre only sources (although at 23 per cent, this detection rate is still quite high). We additionally find that there are more dust clumps that are only associated with water masers (41) than are only associated with methanol masers (13). This suggests that water masers can be present at an even earlier evolutionary stage than methanol masers.

Comparison between the 1.2-mm dust clump properties with and without associated water maser emission, shows (similarly to results of \citet{Breen}) that the water masers are associated with the bigger, more massive sources with higher peak and integrated flux densities. There is a trend whereby the more luminous water masers are associated with 1.2-mm dust clumps with lower H$_2$ number densities than the less luminous water masers. This trend is also seen in the case of both 6.7- and 12.2-GHz methanol masers \citep{Breen10a,Breen12stats}.

Like the 6.7- and 12.2-GHz methanol masers presented by \citet{Breen12stats} we find that there is evidence for both the luminosity and the velocity range of the water masers to increase as the sources evolve. This implies that the gas volume conducive to the maser emission also increases with evolution. The water maser sources show evidence for this trend to `turn over' towards the end of their lifetime, presumably showing a decline in both luminosity and velocity range until the emission ceases.

We have used these water maser observations to test the model for water maser presence towards 1.2-mm dust clumps presented in \citet{Breen}. We find that there is a large number of water maser detections towards dust sources for which the computed probability of water maser presence in greater than 0.01, with a detection rate of 67 per cent towards these sources. However, since the number of clumps where we detect water maser  emission (128) is higher than the number of clumps for which the probability of water maser presence is greater than 0.01 (58) it is clear the the model needs some refinement. The inadequacy of the current model may be (at least partially) attributed to the adoption of the near kinematic distance for sources where distance ambiguities exist, since an assignment of the far distance would result in a larger probability in sources where the water masers are detected. 

We have attempted to create a new model for water maser presence towards 1.2-mm dust clumps, but find that we are severely limited by distance uncertainties. Our analysis shows that the prospects of creating a reliable model for water maser presence within 1.2-mm dust clumps is high when reliable distances can be assigned to the sources. Comparing the success of our model with that of \citet{Palla91} shows that our model has much greater success in predicting water maser detectability.

%a much higher detection rate in the sources that have large calculated probabilities of water maser presence. 

A crude evolutionary implication of our derived model (in conjunction with the original model we present by \citet{Breen}), is that dust clumps with radii equal to 0.97$\pm$0.29 pc (calculated using the difference between the radii threshold implied from each model) have a 50 per cent chance of forming one or more sources that are able to excite water maser emission. Although a simplistic view, this further implies that the lifetime of the dust clump is approximately twice that of water masers.

\section*{Acknowledgments}

We gratefully acknowledge many useful discussions with James Caswell which have increased the quality of the paper. The Australia Telescope Compact Array is part of the Australia Telescope which is funded by the Commonwealth of Australia for operation as a National Facility managed by CSIRO. This research has made use of: NASA's Astrophysics
Data System Abstract Service; the NASA/
IPAC Infrared Science Archive (which is operated by the Jet Propulsion
Laboratory, California Institute of Technology, under contract with
the National Aeronautics and Space Administration); the SIMBAD data base, operated at CDS, Strasbourg,
France; and data products from the GLIMPSE
survey, which is a legacy science program of the {\em Spitzer Space
  Telescope}, funded by the National Aeronautics and Space
Administration.

\end{document}